\documentclass[reprint,superscriptaddress,showpacs]{revtex4-1}
\usepackage[pdftex]{graphicx}
\usepackage{epstopdf}
\usepackage{caption}
\usepackage{multirow}
\usepackage{varioref}
\usepackage{hyperref}
\usepackage{color}

\graphicspath{ {./Figures/} }

\begin{document}

\title[Results and Analysis of LPF Radiation Monitor Data]{Measuring the Galactic Cosmic Ray Flux with the LISA Pathfinder Radiation Monitor}

\def\addressa{European Space Astronomy Centre, European Space Agency, Villanueva de la Ca\~{n}ada, 28692 Madrid, Spain}
\def\addressb{Albert-Einstein-Institut, Max-Planck-Institut f\"ur Gravitationsphysik und Leibniz Universit\"at Hannover, Callinstra{\ss}e 38, 30167 Hannover, Germany}
\def\addressc{APC, Univ Paris Diderot, CNRS/IN2P3, CEA/lrfu, Obs de Paris, Sorbonne Paris Cit\'e, France}
\def\addressd{High Energy Physics Group, Physics Department, Imperial College London, Blackett Laboratory, Prince Consort Road, London, SW7 2BW, UK }
\def\addresse{Dipartimento di Fisica, Universit\`a di Roma ``Tor Vergata'',  and INFN, sezione Roma Tor Vergata, I-00133 Roma, Italy}
\def\addressf{Department of Industrial Engineering, University of Trento, via Sommarive 9, 38123 Trento, and Trento Institute for Fundamental Physics and Application / INFN}
\def\addressh{European Space Technology Centre, European Space Agency, Keplerlaan 1, 2200 AG Noordwijk, The Netherlands}
\def\addressi{Dipartimento di Fisica, Universit\`a di Trento and Trento Institute for Fundamental Physics and Application / INFN, 38123 Povo, Trento, Italy}
\def\addressj{The School of Physics and Astronomy, University of Birmingham, Birmingham, UK}
\def\addressl{Institut f\"ur Geophysik, ETH Z\"urich, Sonneggstrasse 5, CH-8092, Z\"urich, Switzerland}
\def\addressm{The UK Astronomy Technology Centre, Royal Observatory, Edinburgh, Blackford Hill, Edinburgh, EH9 3HJ, UK}
\def\addressn{Institut de Ci\`encies de l'Espai (CSIC-IEEC), Campus UAB, Carrer de Can Magrans s/n, 08193 Cerdanyola del Vall\`es, Spain}
\def\addresso{DISPEA, Universit\`a di Urbino ``Carlo Bo'', Via S. Chiara, 27 61029 Urbino/INFN, Italy}
\def\addressp{European Space Operations Centre, European Space Agency, 64293 Darmstadt, Germany }
\def\addressq{Physik Institut, Universit\"at Z\"urich, Winterthurerstrasse 190, CH-8057 Z\"urich, Switzerland}
\def\addressr{SUPA, Institute for Gravitational Research, School of Physics and Astronomy, University of Glasgow, Glasgow, G12 8QQ, UK}
\def\addresss{Department d'Enginyeria Electr\`onica, Universitat Polit\`ecnica de Catalunya,  08034 Barcelona, Spain}
\def\addresst{Institut d'Estudis Espacials de Catalunya (IEEC), C/ Gran Capit\`a 2-4, 08034 Barcelona, Spain}
\def\addressu{Gravitational Astrophysics Lab, NASA Goddard Space Flight Center, 8800 Greenbelt Road, Greenbelt, MD 20771 USA}
\def\addressbb{Department of Physics, University of Florida, 2001 Museum Rd, Gainesville, FL 32603, USA}
\def\addresscc{Dipartimento di Scienze Fisiche e Chimiche, Universit\`a degli Studi dell’Aquila, Via Vetoio, Coppito, 67100 L’Aquila, Italy}
\def\addressdd{Sezione INFN di Firenze, Via G. Sansone n. 1, 50019 Sesto Fiorentino (Firenze), Italy}

\author{M~Armano}\affiliation{\addressa}
\author{H~Audley}\affiliation{\addressb}
\author{J~Baird}\affiliation{\addressd}
\author{P~Binetruy}\thanks{Deceased 30 March 2017}\affiliation{\addressc}
\author{M~Born}\affiliation{\addressb}
\author{D~Bortoluzzi}\affiliation{\addressf}
\author{E~Castelli}\affiliation{\addressi}
\author{A~Cavalleri}\affiliation{\addressi}
\author{A~Cesarini}\affiliation{\addresso}
\author{A\,M~Cruise}\affiliation{\addressj}
\author{K~Danzmann}\affiliation{\addressb}
\author{M~de Deus Silva}\affiliation{\addressa}
\author{I~Diepholz}\affiliation{\addressb}
\author{G~Dixon}\affiliation{\addressj}
\author{R~Dolesi}\affiliation{\addressi}
\author{L~Ferraioli}\affiliation{\addressl}
\author{V~Ferroni}\affiliation{\addressi}
\author{N~Finetti}\affiliation{\addresscc}\affiliation{\addressdd}
\author{E\,D~Fitzsimons}\affiliation{\addressm}
\author{M~Freschi}\affiliation{\addressa}
\author{L~Gesa}\affiliation{\addressn}
\author{F~Gibert}\affiliation{\addressi}
\author{D~Giardini}\affiliation{\addressl}
\author{R~Giusteri}\affiliation{\addressi}
\author{C~Grimani}\affiliation{\addresso}
\author{J~Grzymisch}\affiliation{\addressh}
\author{I~Harrison}\affiliation{\addressp}
\author{G~Heinzel}\affiliation{\addressb}
\author{M~Hewitson}\affiliation{\addressb}
\author{D~Hollington}\email[Corresponding author:]{d.hollington07@imperial.ac.uk}\affiliation{\addressd}
\author{D~Hoyland}\affiliation{\addressj}
\author{M~Hueller}\affiliation{\addressi}
\author{H~Inchausp\'e}\affiliation{\addressc}
\author{O~Jennrich}\affiliation{\addressh}
\author{P~Jetzer}\affiliation{\addressq}
\author{N~Karnesis}\affiliation{\addressb}
\author{B~Kaune}\affiliation{\addressb}
\author{N~Korsakova}\affiliation{\addressr}
\author{C\,J~Killow}\affiliation{\addressr}
\author{J\,A~Lobo}\thanks{Deceased 30 September 2012}\affiliation{\addressn}
\author{I~Lloro}\affiliation{\addressn}
\author{L~Liu}\affiliation{\addressi}
\author{J\,P~L\'opez-Zaragoza}\affiliation{\addressn}
\author{R~Maarschalkerweerd}\affiliation{\addressp}
\author{D~Mance}\affiliation{\addressl}
\author{N~Meshskar}\affiliation{\addressl}
\author{V~Mart\'{i}n}\affiliation{\addressn}
\author{L~Martin-Polo}\affiliation{\addressa}
\author{J~Martino}\affiliation{\addressc}
\author{F~Martin-Porqueras}\affiliation{\addressa}
\author{I~Mateos}\affiliation{\addressn}
\author{P\,W~McNamara}\affiliation{\addressh}
\author{J~Mendes}\affiliation{\addressp}
\author{L~Mendes}\affiliation{\addressa}
\author{M~Nofrarias}\affiliation{\addressn}
\author{S~Paczkowski}\affiliation{\addressb}
\author{M~Perreur-Lloyd}\affiliation{\addressr}
\author{A~Petiteau}\affiliation{\addressc}
\author{P~Pivato}\affiliation{\addressi}
\author{E~Plagnol}\affiliation{\addressc}
\author{J~Ramos-Castro}\affiliation{\addresss}
\author{J~Reiche}\affiliation{\addressb}
\author{D\,I~Robertson}\affiliation{\addressr}
\author{F~Rivas}\affiliation{\addressn}
\author{G~Russano}\affiliation{\addressi}
\author{J~Slutsky}\affiliation{\addressu}
\author{C\,F~Sopuerta}\affiliation{\addressn}
\author{T~Sumner}\affiliation{\addressd}
\author{D~Texier}\affiliation{\addressa}
\author{J\,I~Thorpe}\affiliation{\addressu}
\author{D~Vetrugno}\affiliation{\addressi}
\author{S~Vitale}\affiliation{\addressi}
\author{G~Wanner}\affiliation{\addressb}
\author{H~Ward}\affiliation{\addressr}
\author{P~Wass}\affiliation{\addressd}\affiliation{\addressbb}
\author{W\,J~Weber}\affiliation{\addressi}
\author{L~Wissel}\affiliation{\addressb}
\author{A~Wittchen}\affiliation{\addressb}
\author{P~Zweifel}\affiliation{\addressl}

\begin{abstract}

Test mass charging caused by cosmic rays will be a significant source of acceleration noise for space-based gravitational wave detectors like LISA. Operating between December 2015 and July 2017, the technology demonstration mission LISA Pathfinder included a bespoke monitor to help characterise the relationship between test mass charging and the local radiation environment. The radiation monitor made \textit{in situ} measurements of the cosmic ray flux while also providing information about its energy spectrum. We describe the monitor and present measurements which show a gradual 40\% increase in count rate coinciding with the declining phase of the solar cycle. Modulations of up to 10\% were also observed with periods of 13 and 26 days that are associated with co-rotating interaction regions and heliospheric current sheet crossings. These variations in the flux above the monitor detection threshold ($\approx70\,\textup{MeV}$) are shown to be coherent with measurements made by the IREM monitor on-board the Earth orbiting INTEGRAL spacecraft. Finally we use the measured deposited energy spectra, in combination with a GEANT4 model, to estimate the galactic cosmic ray differential energy spectrum over the course of the mission.

\end{abstract}

\maketitle


\section{Introduction}

Launched in December 2015 and operated until mission completion in July 2017, LISA Pathfinder (LPF) was a European Space Agency mission that successfully demonstrated the feasibility of building a future space-based gravitational wave observatory \cite{Armano2016}. As currently envisaged, an observatory like the proposed Laser Interferometer Space Antenna (LISA) will involve placing test masses on-board distant spacecraft in near perfect free-fall while using laser interferometry to measure their relative acceleration \cite{Amaro2017}. At low frequencies a variety of local forces can produce spurious acceleration noise and limit the sensitivity of the detector. One such disturbance results from the noisy accumulation of charge on the free-floating test masses due to the high-energy ionising radiation present in the space environment. Such charging has been shown to be a significant source of acceleration noise at frequencies below $1\,\textup{mHz}$ \cite{Armano2017}.

The Pathfinder orbit around the L1 Lagrange point placed it outside Earth's protective magnetosphere and exposed it to the interplanetary charged particle environment. To contribute to test mass charging particles needed to be of an energy sufficient to penetrate the spacecraft and outer housing with simulation predicting this cut-off to be around $100\,\textup{MeV}$ for protons \cite{Araujo2005, Wass2005}.

The test mass charging rate itself is dependant on both the flux and energy spectrum of the incident radiation. However, if either of these are modulated, for example by fluctuations in the Interplanetary Magnetic Field (IMF), it can lead to excess charging noise above the expected flat Poissonian spectrum. This outcome is fairly intuitive in terms of the incident flux while for a modulated energy spectrum one needs to consider that due to higher charge multiplicity, charging from high-energy particles is more noisy than charging from lower energy particles \cite{Araujo2005}.

In order to better understand the relationship between test mass charging and the local radiation environment the Pathfinder payload included a bespoke radiation monitor. Its purpose was to make \textit{in situ} measurements of the flux while also providing information about the energy spectrum of the incident radiation.


\section{The Interplanetary Charged Particle Environment}
\label{sec:LIS}

Above the $100\,\textup{MeV}$ boundary relevant for Pathfinder there are two main sources of interplanetary charged particle that dominate test mass charging. Forming a permanent background, the first originate from outside the solar system and are referred to as Galactic Cosmic Rays (GCRs). These primarily consist of protons but also contain a significant fraction of helium nuclei ($\alpha$-particles) as well as a small fraction of heavier nuclei. The precise ratios vary at lower energies but above several GeV approximately 80\% of primary nucleons are free protons and about 70\% of the rest are bound in helium nuclei \cite{Astro2004}. At lower energies, measurements within the heliosphere have shown that the GCR background fluctuates over various time-scales with changes over the 11-year solar cycle being most significant.

A common way of understanding these modulations is to assume an isotropic, steady-state flux at the heliosphere boundary, referred to as the local interstellar spectrum (LIS). This flux of particles then interacts with the solar wind and IMF as it penetrates deeper within the heliosphere with variations in the heliospheric properties therefore leading to the observed temporal changes in the GCR spectrum. The change in flux is strongly dependant on the energy of the particle with those at around $100\,\textup{MeV}$ varying by orders of magnitude while the flux of particles above about $10\,\textup{GeV}$ being almost constant. 

\begin{figure}[h]
\centering
\includegraphics[width=0.45\textwidth]{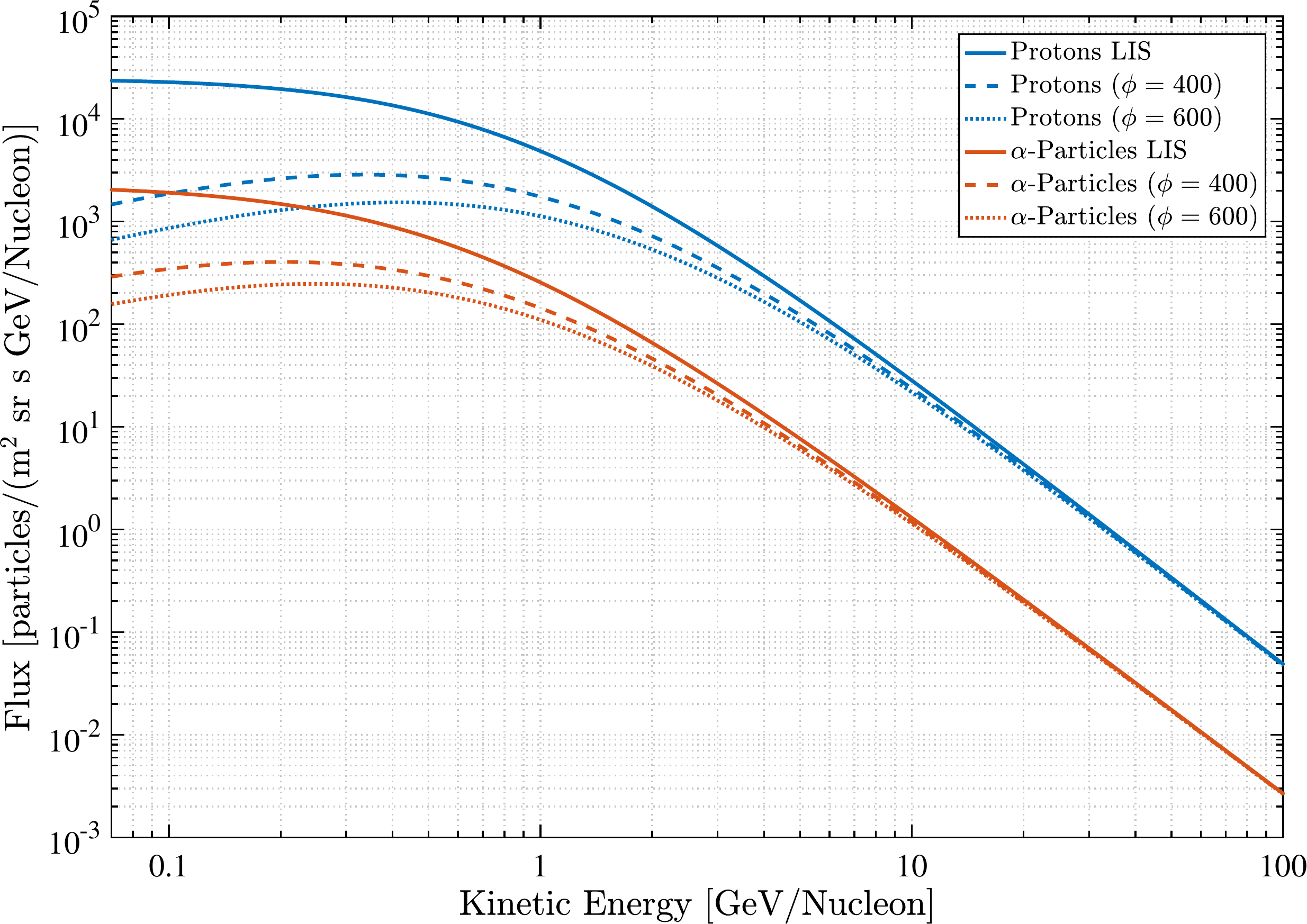}
\caption[The expected proton flux.]{\label{fig:predictedFlux} The differential energy spectra for both protons and $\alpha$-particles, where $\phi$ takes values that were at the extremes of those applicable for Pathfinder. With $\phi=400$ the integrated flux above $70\,\textup{MeV}$ for protons is 0.47 cm$^{-2}$ sr$^{-1}$ s$^{-1}$ while for $\phi=600$ it is 0.32 cm$^{-2}$ sr$^{-1}$ s$^{-1}$. The unattenuated LIS for both are also shown for comparison.}
\end{figure}

Given the difficulty in measuring it directly, the LIS has historically been inferred from measurements made within the heliosphere although in recent years Voyager 1 has begun providing \textit{in situ} measurements \cite{Stone2013}. Several expressions that describe the LIS for protons and $\alpha$-particles can be found within the literature but throughout this paper we will use those described by Bisschoff et al. \cite{Bisschoff2016} which are based on combined Voyager 1 and PAMELA measurements:

\begin{equation}
J_{LIS_p}(T) = 3719.0\frac{1}{\beta^{2}}T^{1.03}\left(\frac{T^{1.21} + 0.77^{1.21}}{1+0.77^{1.21}}\right)^{-3.18}
\label{eq:LIS_p}
\end{equation}

\begin{equation}
J_{LIS_\alpha}(T) = 195.4\frac{1}{\beta^{2}}T^{1.02}\left(\frac{T^{1.19} + 0.60^{1.19}}{1+0.60^{1.19}}\right)^{-3.15}
\label{eq:LIS_alpha}
\end{equation}

\noindent where $T$ is the kinetic energy (GeV/nucleon), $\beta=v/c$ is the particles velocity relative to the speed of light and $J_{LIS}(T)$ is in units of (particles/(m$^{2}$ sr s GeV/nucleon)). The time dependant differential intensity $J_{i}$ of nucleus of type $i$ at 1 AU can then be parametrised by the ``force-field" approximation \cite{Gleeson1968, Usoskin2005, Usoskin2011}:

\begin{equation}
J_{i}(T,\phi) = J_{LIS,i}(T+\Phi)\frac{(T)(T+2T_{r})}{(T+\Phi)(T+\Phi+2T_{r})}
\label{eq:LIS(phi)}
\end{equation}

\noindent where $T_{r}$ is the rest mass (GeV/nucleon), $\Phi=(Ze\cdot10^{-3}/A)\phi$, $Z$ is the atomic number, $e$ is the elementary charge, $A$ is the mass number and $\phi$ is the modulation potential (MV). The parameter $\phi$ has limited physical meaning and is implicitly dependant on the fixed shape of the LIS chosen. However, it offers a simple way of describing both the proton and $\alpha$-particle differential energy spectra incident on the Pathfinder spacecraft using just a single parameter. Figure \ref{fig:predictedFlux} illustrates the relationship between the above LIS and $\phi$, where $\phi$ takes values of 400 \& 600 MV which were the values at either extreme of the fluxes observed during the Pathfinder mission (as we shall see in Sec. \ref{sec:modelling}).

The second source of charged particles above $100\,\textup{MeV}$ are those shock accelerated near the Sun during coronal mass ejections, referred to as Solar Energetic Particles (SEPs). These transient solar eruptions are more frequent around solar maximum, with on average a few a year energetic enough to enhance test mass charging, but less than one a year at solar minimum. Their properties are event specific but can temporarily increase the proton flux by several orders of magnitude for durations spanning hours up to several days \cite{Wass2005, Adriani2011, Grimani2014}. Given the Pathfinder mission was to be operational for around 18 months during the declining phase of the solar cycle, pre-flight predictions were for at most one solar event energetic enough to enhance test mass charging \cite{Grimani2012}.


\section{Hardware}

The Pathfinder radiation monitor was developed and built by a Barcelona based consortium which included NTE-SENER, the Institut de Ci\`encies de l'Espai (IEEC-CSIC) and the Institut de F\'isica d'Altes Energies (IFAE) \cite{Canizares2011} with design and testing being aided by software simulations carried out at Imperial College London \cite{Wass2006}. Based on a simple design, it consisted of two silicon PIN diodes in a telescopic configuration and recorded both a singles count as well as the deposited energy spectrum of coincident events.

\begin{figure}[]
\centering
\includegraphics[height=0.16\textheight]{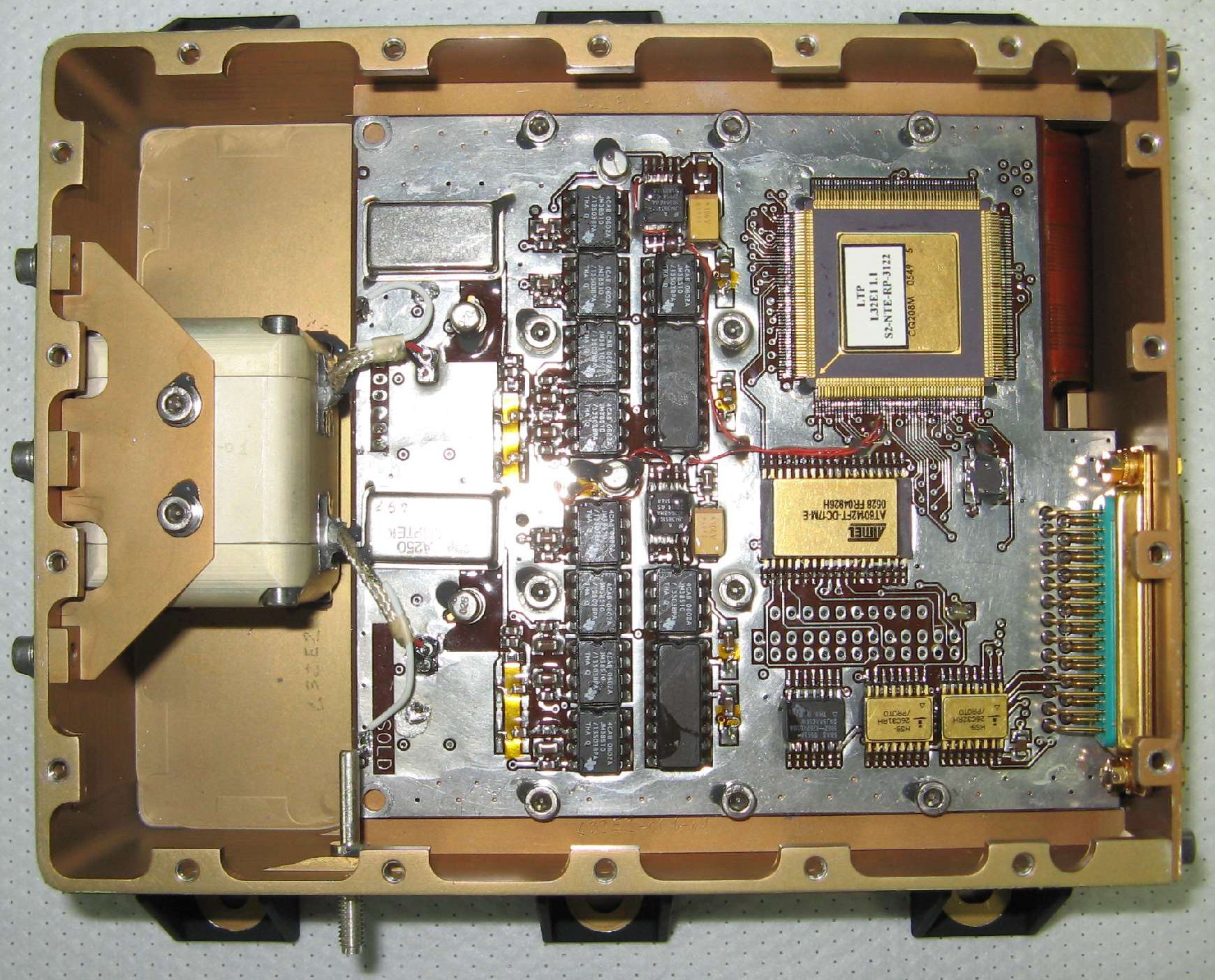}
\hfil
\includegraphics[height=0.16\textheight]{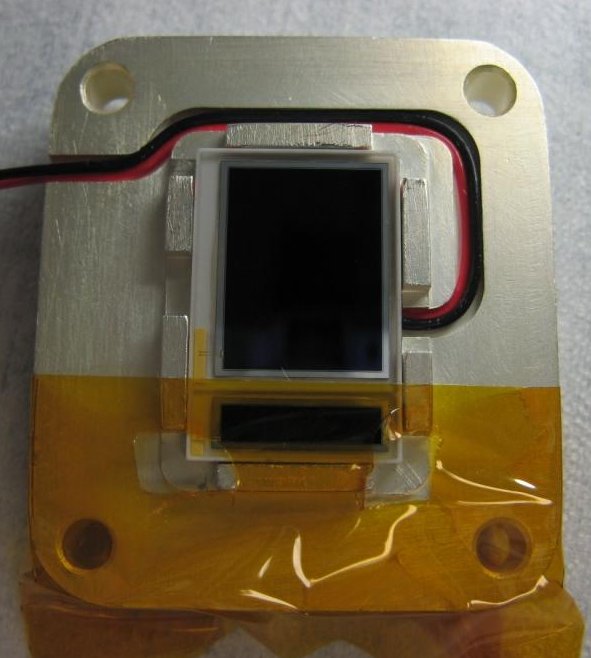}
\caption[The LPF radiation monitor.]{\label{fig:RMphotos} Left: A top-down photograph of the radiation monitor, with the lid removed. The cubic copper shield can be seen on the left, next to the unit's electronics. Right: A production photograph of one of the flight PIN diodes mounted on the wall of its copper shield. Note only the larger black region was sensitive and the \textit{Kapton} tape used to temporarily hold the diode in place was removed prior to integration.}
\end{figure}

The individual diode packages were Hamamatsu dual PIN photodiodes (S8576-01) and were originally flight-spares from the Fermi Gamma-ray Space Telescope (FGST - formerly called the Gamma-ray Large Area Space Telescope or GLAST) which were re-purposed for use on Pathfinder. Each had a $320\,\mu\textup{m}$ thick sensitive area of $10.5\times14.0\,\textup{mm}^{2}$ and were mounted $20.0\,\textup{mm}$ apart within a hollow copper cuboid. This copper shield was designed (with some margin) to prevent protons less than $70\,\textup{MeV}$ being detected as early modelling had found that only protons above $\approx100\,\textup{MeV}$ were energetic enough to reach the test masses and contribute to charging. The shield's effectiveness was verified by on ground testing \cite{Mateos2012} and had $6.4\,\textup{mm}$ thick walls, rounded corners, measured $43.4\times40.8\times36.1\,\textup{mm}^{3}$ and along with the associated electronics, was housed in the aluminium flight box shown in Figure \ref{fig:RMphotos}.

The radiation monitor was mounted on an internal side wall within the spacecraft, with the telescopic arrangement pointed towards the Sun throughout the mission being aligned along a normal to the external solar panels. In terms of the total energy deposited in a single diode the detection threshold was nominally set at $60.7\,\textup{keV}$, which was a compromise between being above the electronics noise while not losing a significant number of actual counts. The radiation monitor kept a count of the combined number of events above this threshold in both the front and rear diodes, so called singles events. If both were triggered within $525\,\textup{ns}$ of each other the energy deposited within the rear diode was recorded, a coincident event. Note that due to the design of the radiation monitor electronics a coincident event did not increment the singles count. The number of singles events in a 15 second period was returned in telemetry while a deposited energy spectrum was returned every 600 seconds. Its sensitive range was from $0\,\textup{to}\,5\,\textup{MeV}$, which, with 1024 linear bins, gave a deposited energy resolution of approximately $4.88\,\textup{keV}$.


\section{Radiation Monitor Model}

To aid the design and testing of the Pathfinder radiation monitor a detailed two-part software model was developed over a number of years alongside the hardware itself. The first part was written within the GEANT4 framework which uses Monte-Carlo techniques to accurately simulate the passage of particles through matter \cite{Agostinelli2003, Allison2006, Allison2016}. Given a source of particles, be that a collimated proton beam during testing or a spectrum of isotropically emitted protons to simulate the GCR background, our GEANT4 model outputs a file containing the energy deposited in each diode by each incident particle. The second part of the radiation monitor model was written in MATLAB and takes the deposited energy file as an input, adds a parametrised level of electronics noise to each event, applies the monitor's detection threshold and outputs a simulated singles count rate as well as a deposited energy spectrum. Over the years the complete radiation monitor model has been through three iterations.

The first version was used to optimise the design of the copper shield and verify the performance of a prototype monitor during high-energy proton beam tests carried out at the Paul Scherrer Institute (PSI) in Switzerland  \cite{Wass2006, Wass2007}. This model simply contained the diodes, ceramic mountings, an internal printed circuit board as well as the surrounding copper shield. A Gaussian electronics noise with a mean of zero and a pessimistic $\sigma=12.7\,\textup{keV}$ was assumed with a hard threshold cut at $50\,\textup{keV}$. Alongside design and testing support, the simulation was also used to provide expected in-flight count rates due to the GCR flux. It predicted a singles count of $3.3\,\textup{counts s}^{-1}$ and a coincident count of $0.17\,\textup{counts s}^{-1}$ during solar maximum with the rates increasing to $7.2\,\textup{counts s}^{-1}$ and $0.37\,\textup{counts s}^{-1}$ during solar minimum \cite{Wass2006}.

\begin{figure}[]
\centering
\includegraphics[height=0.16\textheight]{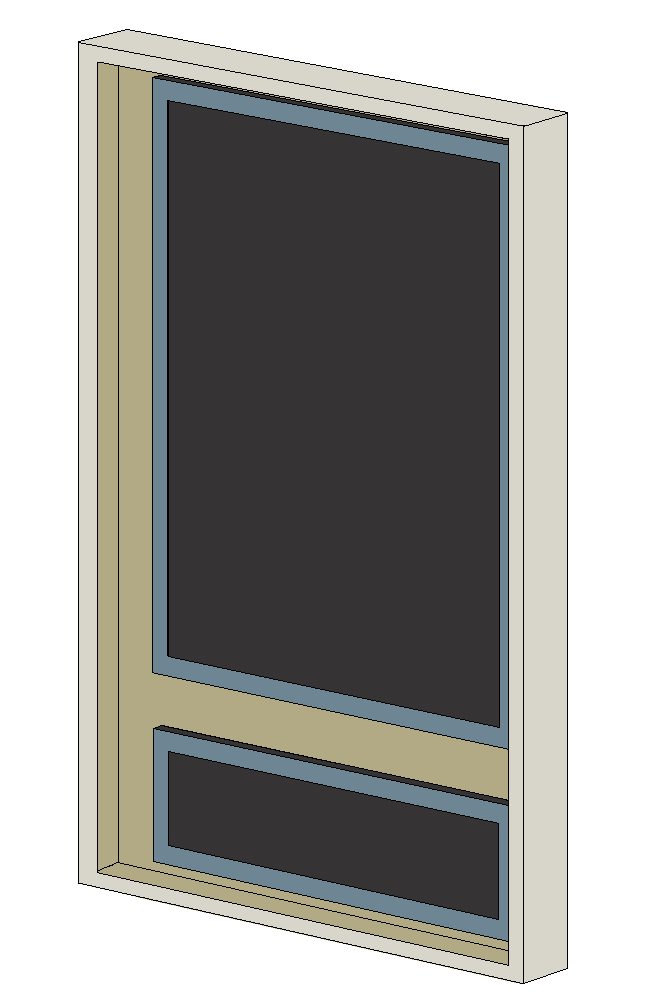}
\hfil
\includegraphics[height=0.16\textheight]{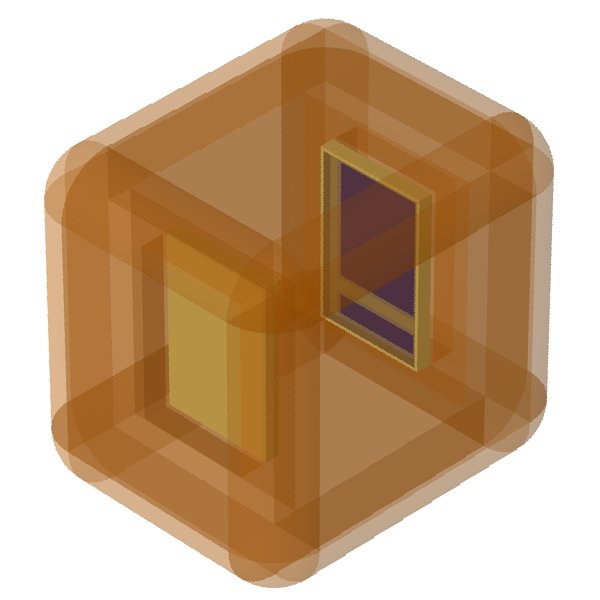}
\hfil
\includegraphics[height=0.16\textheight]{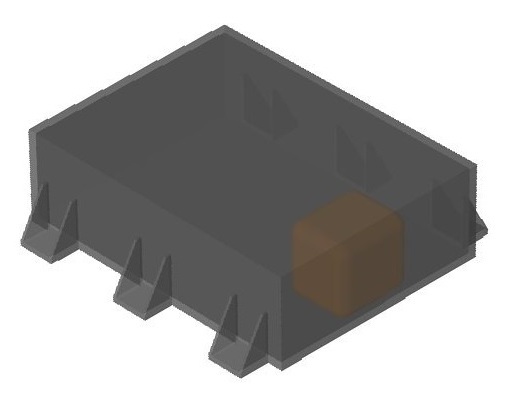}
\caption[The GEANT4 radiation monitor model.]{\label{fig:RMsim} A visualisation of the simulated flight-model radiation monitor. Clockwise from top left: an individual diode, two diodes mounted in their telescopic configuration within the copper shield and finally the enclosing aluminium flight box.}
\end{figure}

The flight radiation monitor was also tested at the PSI proton beam facility, \cite{Mateos2012}. This led to a more refined simulated model of the monitor being developed which included a more detailed representation of the diodes and added the aluminium flight box which housed the copper shield alongside the associated electronics \cite{Hollington2011}, see Figure \ref{fig:RMsim}. Using the simulation to model the beam test results confirmed that the sensitive regions of the two diodes were in excellent angular alignment ($+0.3^{\circ}\pm0.3^{\circ}$) but had a small linear offset between them ($1.0\pm0.2\,\textup{mm}$) due to the way they were mounted. In addition it was realised that the sensitive regions were $320\,\mu\textup{m}$ thick (rather than the $300\,\mu\textup{m}$ value assumed previously), with this fact later confirmed by the discovery in the literature of an x-ray cross-section of the FGST (formerly named GLAST) diodes \cite{Serma2002}. The electronics noise and threshold components remained unchanged.

The third and final iteration of the radiation monitor added a $1.42\,\textup{mm}$ thick aluminium spherical shell around the flight box to approximate the spacecraft. The main advancement of the model came from several in-flight calibration measurements that provided a better understanding of the electronics noise and threshold properties as well as how they evolved over the duration of the mission. This will be discussed in detail within Sec. \ref{sec:modelling}.

\section{In-Flight Measurements}

Following the successful launch of Pathfinder on the $3^{\textup{rd}}$ December 2015, the spacecraft reached its final orbit around the L1 Lagrange point in February 2016. During transit the radiation monitor was temporarily turned on for 10 days from the $11^{\textup{th}}$ January before being turned off until the $26^{\textup{th}}$ January. Following a 5 day commissioning period where the detection threshold was adjusted, the radiation monitor then collected data continuously from $15^{\textup{th}}$ February 2016 until it was turned off on the $17^{\textup{th}}$ July 2017 at the mission's end. Over this period there were a few brief outages due to planned resets of the system wide electronics unrelated to radiation monitor performance (ten's of minutes) and a repeat of the threshold calibration in April 2017 that lasted 7 days. With these exceptions the now publicly available (\url{http://lpf.esac.esa.int/lpfsa/}) radiation monitor data presented here spans the entirety of the Pathfinder mission.

\begin{figure*}[]
\centering
\includegraphics[width=0.8\textwidth]{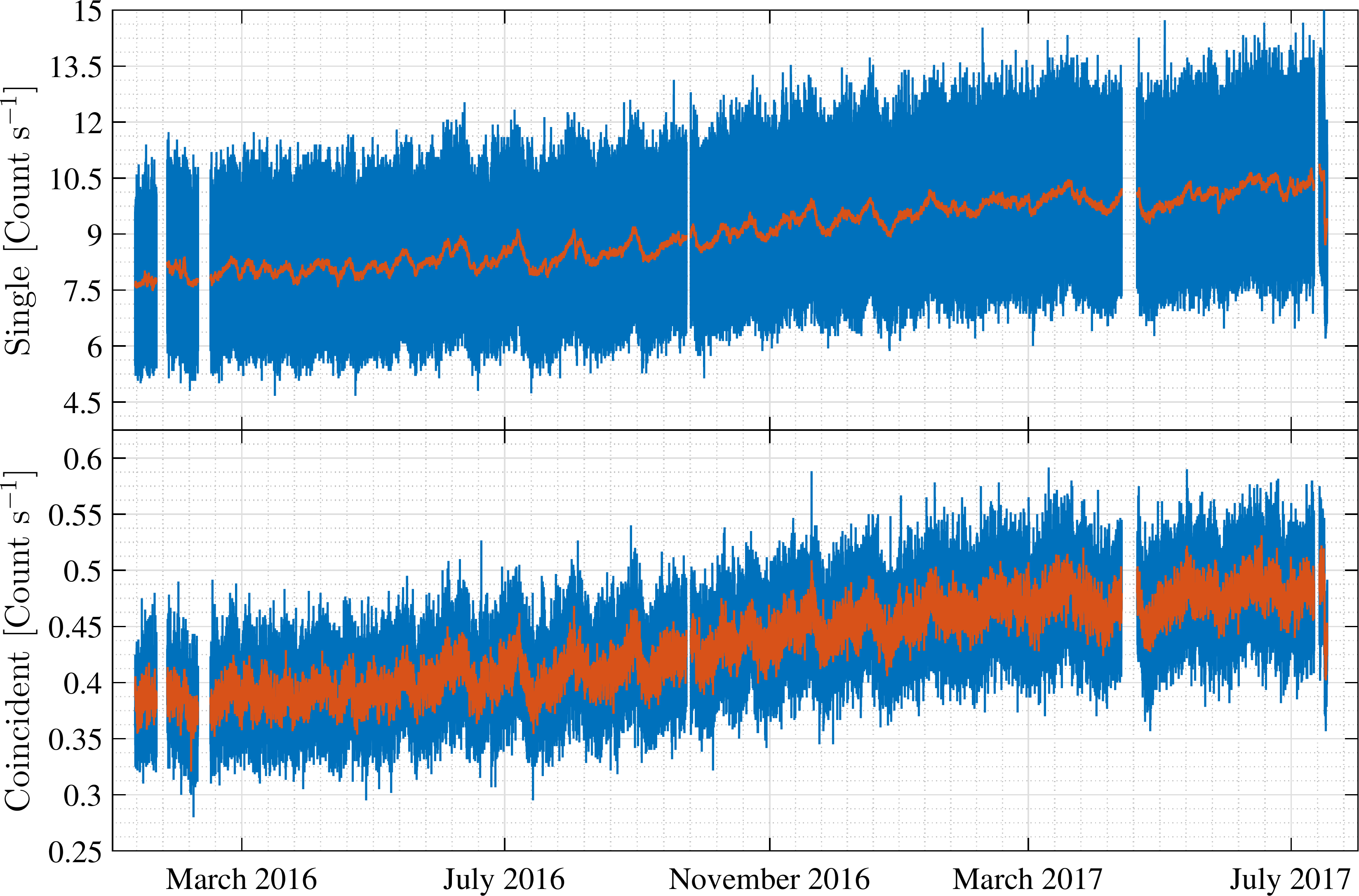}
\setlength\lineskip{10pt}
\includegraphics[width=0.48\textwidth]{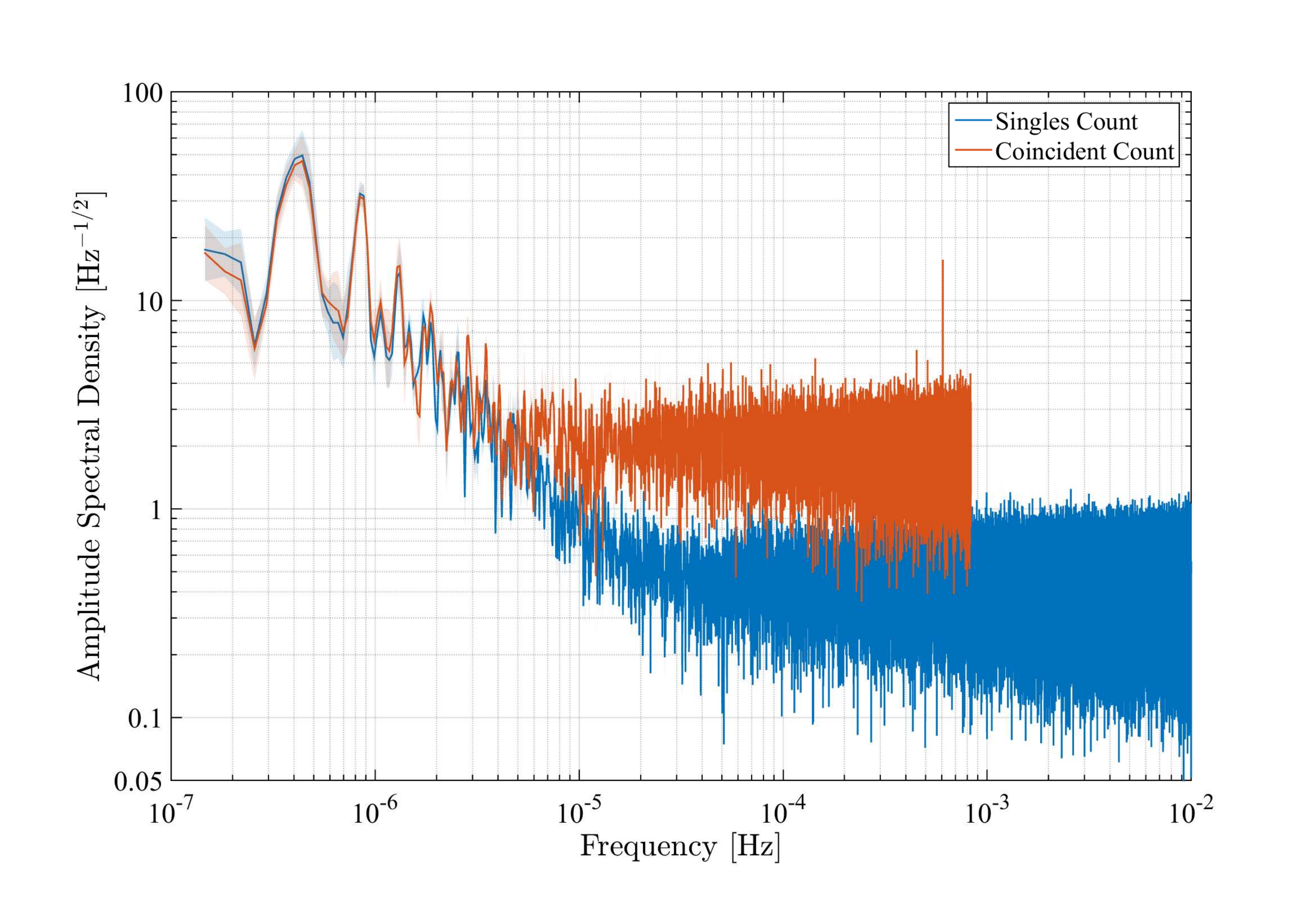}
\centering
\includegraphics[width=0.45\textwidth]{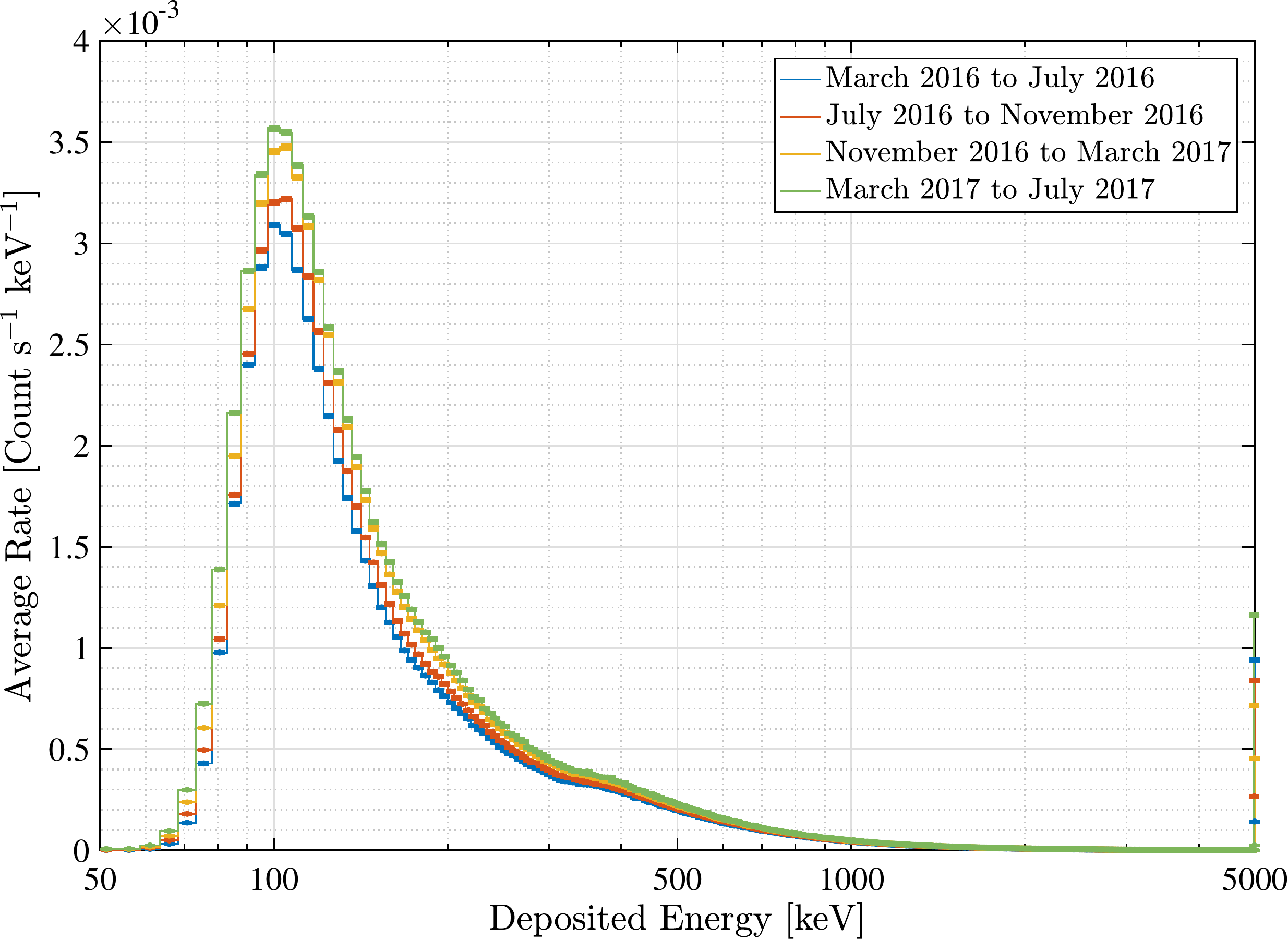}
\caption[The LPF RM singles and coincident count.]{\label{fig:countsPlotFull} Top: The raw singles ($1/15\,\textup{Hz}$) and coincident ($1/600\,\textup{Hz}$) count data spanning the whole mission with hourly averaged data overlaid. Bottom Left: The amplitude spectral density of the singles and coincident channels over the entire mission. Bottom Right: Averaged deposited energy spectra, each containing data measured over four months.}
\end{figure*}

Figure \ref{fig:countsPlotFull} shows both the singles and coincident event count rates as measured over the course of the mission. In both cases the data are shown at the original sampling frequency (blue) and with hourly averaged data overlaid (red). As expected, the coincident count rate is noisier than the singles count rate due to the lower number of events and reassuringly both are highly correlated. Over the course of the mission both rates showed a general trend upwards with the mean singles count rate increasing from  $7.6\,\textup{to}\,10.6\,\textup{counts s}^{-1}$ and the mean coincident count rate from $0.38\,\textup{to}\,0.51\,\textup{counts s}^{-1}$. However, in both channels there was a clear underlying oscillation with a period that varied between 13 and 26 days with a peak to peak variation of around 10\% in the singles count. The amplitude spectral density (also shown in Figure \ref{fig:countsPlotFull}) of both the singles and coincident count has a peak corresponding to the 26 day oscillation as well as its harmonics. We shall return to the nature of this oscillation in the next section.

Figure \ref{fig:countsPlotFull} also shows the average deposited energy spectra integrated over four month periods, after correcting for electronics drift (described in Sec. \ref{sec:modelling}). We see little to no change in the relative shape or position of the spectra, merely a scaling proportional to the total coincident count rate. This ability to record the deposited energy spectrum of incident particles was primarily included to allow differentiation between the energy composition of the permanent GCR flux and that from any transient SEP events. During the mission, as it happened, no SEP events occurred with particle energies high enough to reach the test masses and therefore enhance the charging rate. However, there were several minor SEP events which were observed by the Earth orbiting Geostationary Operational Environmental Satellites (GOES) which showed flux enhancements for protons with energies up to $40\,\textup{MeV}$.

One occurred on the $16^{\textup{th}}$ March 2016 while another on $16^{\textup{th}}$ May 2016, with increased activity lasting 28 and 37 hours respectively. Although the Pathfinder radiation monitor should have been insensitive to such low energy events we checked for enhancements in the singles count rate nonetheless. No change was observed during the second event, though there was a minor increase of around 2\% during the first. However, this could have been due to an increased GCR count resulting from the turbulent IMF during this time rather than a direct detection of the SEP event itself. 

\begin{figure}[]
\centering
\includegraphics[width=0.45\textwidth]{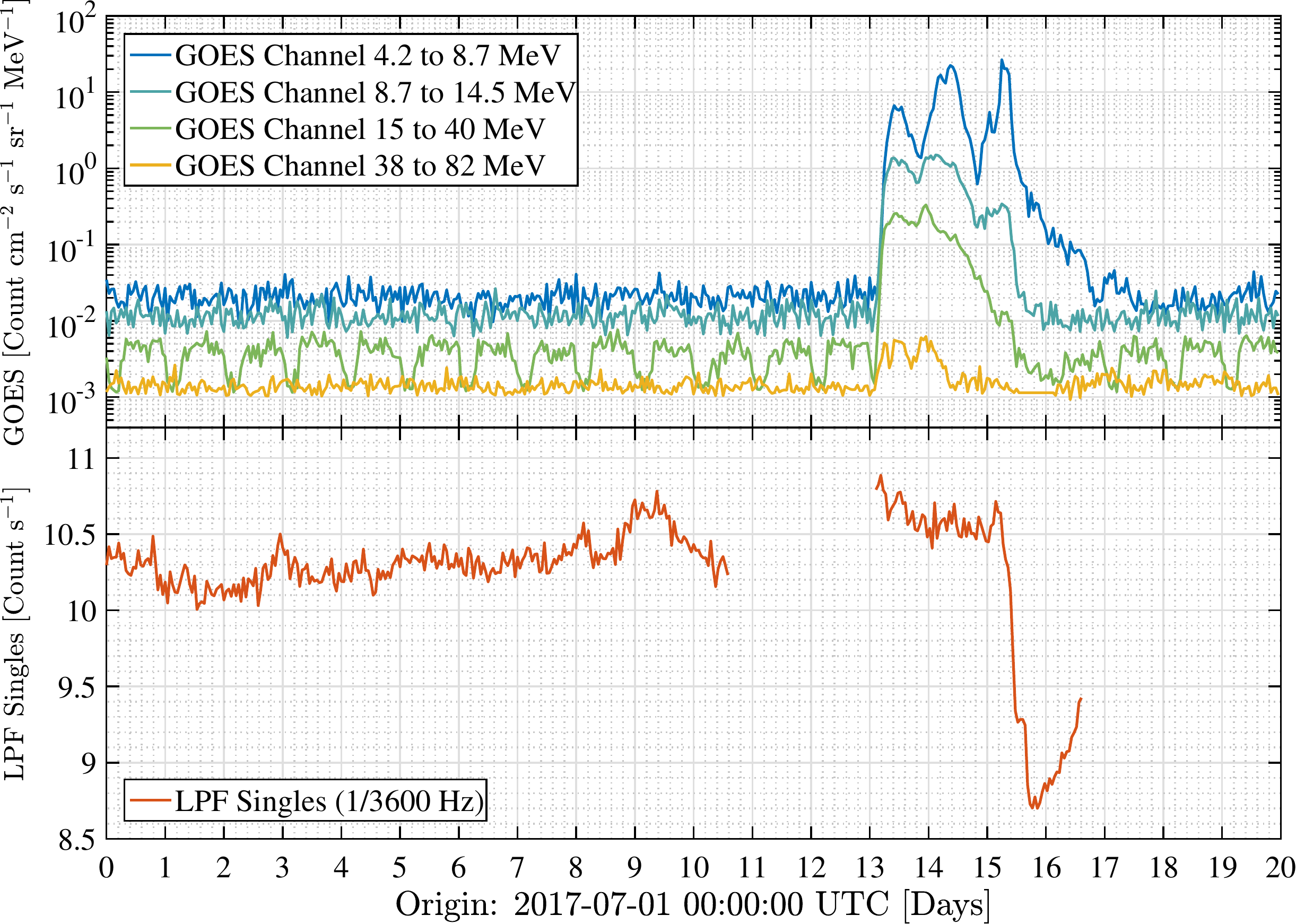}
\caption[Forbush Decrease.]{\label{fig:forbushDecrease} Top: The low energy proton channels as measured by the GOES-15 satellite. A relatively minor SEP event was observed which began on the $14^{\textup{th}}$ July 2017 and persisted for several days. Bottom: The hourly averaged singles count channel from the Pathfinder radiation monitor. No enhancement was observed but a $\sim18\%$ Forbush decrease was seen as the event subsided, just prior to the Pathfinder mission coming to an end. Note that the 2.5 day loss of data before the SEP event was coincidental and due to a planned but unrelated high temperature experiment on-board Pathfinder.}
\end{figure}

The largest SEP event occurred just days before the final turn-off of the monitor. It was detected by the GOES satellites on the $14^{\textup{th}}$ July 2017 and persisted for several days. Like the other events, GOES did not detect any flux enhancements above $100\,\textup{MeV}$ and as expected no increase in count rate was observed by the Pathfinder monitor. However, the Pathfinder monitor did observe a significant Forbush decrease \cite{Forbush1937}, shown in Figure \ref{fig:forbushDecrease}. An approximate $18\%$ fall in the singles and coincident count was seen in just 13 hours, significant when compared to the gradual $\sim40\%$ increase observed over the previous 18 months of observation. Unfortunately, due to the scheduled end of the Pathfinder mission we only managed to observe the partial recovery of the GCR flux.

\subsection{Comparison with other Instruments}
\label{sec:comparison}

In order to make comparisons with data from other instruments we shall in this section restrict ourselves to examining a four month period of data (March 2016 to July 2016) which encompassed the first phase of Pathfinder mission operations. First, in order to confirm that the observed 13 and 26 day oscillations in the count rate were due to physical changes in the radiation environment, rather than a hardware issue, we compared the measurements to those made on-board the Earth orbiting INTEGRAL spacecraft. 

The INTEGRAL Radiation Environment Monitor (IREM) utilises three silicon detectors and is designed to measure protons with energies $E>10\,\textup{MeV}$ and electrons $E>0.5\,\textup{MeV}$. The spacecraft itself is in a highly eccentric Earth orbit with a period of three days, resulting in it spending 7 to 10 hours every orbit within Earth's radiation belts, which then dominate the IREM signal. However, outside of these times the IREM effectively measures GCRs and of particular interest here is the TC2 channel which is sensitive to protons with $E>49\,\textup{MeV}$ and electrons $E>2.8\,\textup{MeV}$. Combining prior knowledge of the GCR spectrum and the instruments derived response function \cite{Sandberg2012}, the TC2 channel, like the Pathfinder monitor, should be dominated by protons with $E>70\,\textup{MeV}$.

Figure \ref{fig:iremComparison} shows a comparison between the Level-0 IREM TC2 channel (radiation belt crossings removed) with the singles count measured by the Pathfinder monitor during the first phase of science operations. As one can seen, there is excellent agreement between the two measurements albeit with the absolute Pathfinder count being about 2.5 times higher than the IREM due to its larger sensitive area. Although the IREM data is noisier and inevitably contains gaps (due to radiation belt crossings) the equivalence with the Pathfinder data opens some interesting future possibilities. For example, the IREM dataset goes back to October 2002 providing a measurement of the proton flux relevant to test mass charging over an entire solar cycle and also having recorded many SEP events. Regardless, the IREM data not only shows that the 13 and 26 day oscillations are real but they are likely due to large scale variations in the heliosphere given they were detected on spacecraft separated by around $1.5\times10^{6}\,\textup{km}$.

\subsection{GCR Modulation}
\label{sec:GCRmodulation}

The heliosphere is a complex dynamical system formed by the continuous emission of charged particles from the upper atmosphere of the Sun and its time varying nature can act to modulate the observed GCR flux in a number of ways, both locally and on larger scales. On the scale of the solar system this is most noticeable over the 11-year solar cycle where GCR flux increases at solar minimum when the overall magnitude of the IMF is weaker and decreases at solar maximum when it is stronger. On smaller scales, the effects of localised activity on the Sun can sweep out into the Solar System with a particular directionality. The resulting changes in certain regions of the IMF can then influence the received GCR flux locally. If such solar events persist for long enough they can appear to come and go due to solar rotation. Given that the synodic rotation period (the time taken for a fixed point on the solar equator to rotate to the same apparent position as viewed from Earth) is around 26 days one would expect GCR modulations on similar time-scales.

\begin{figure}[]
\centering
\includegraphics[width=0.45\textwidth]{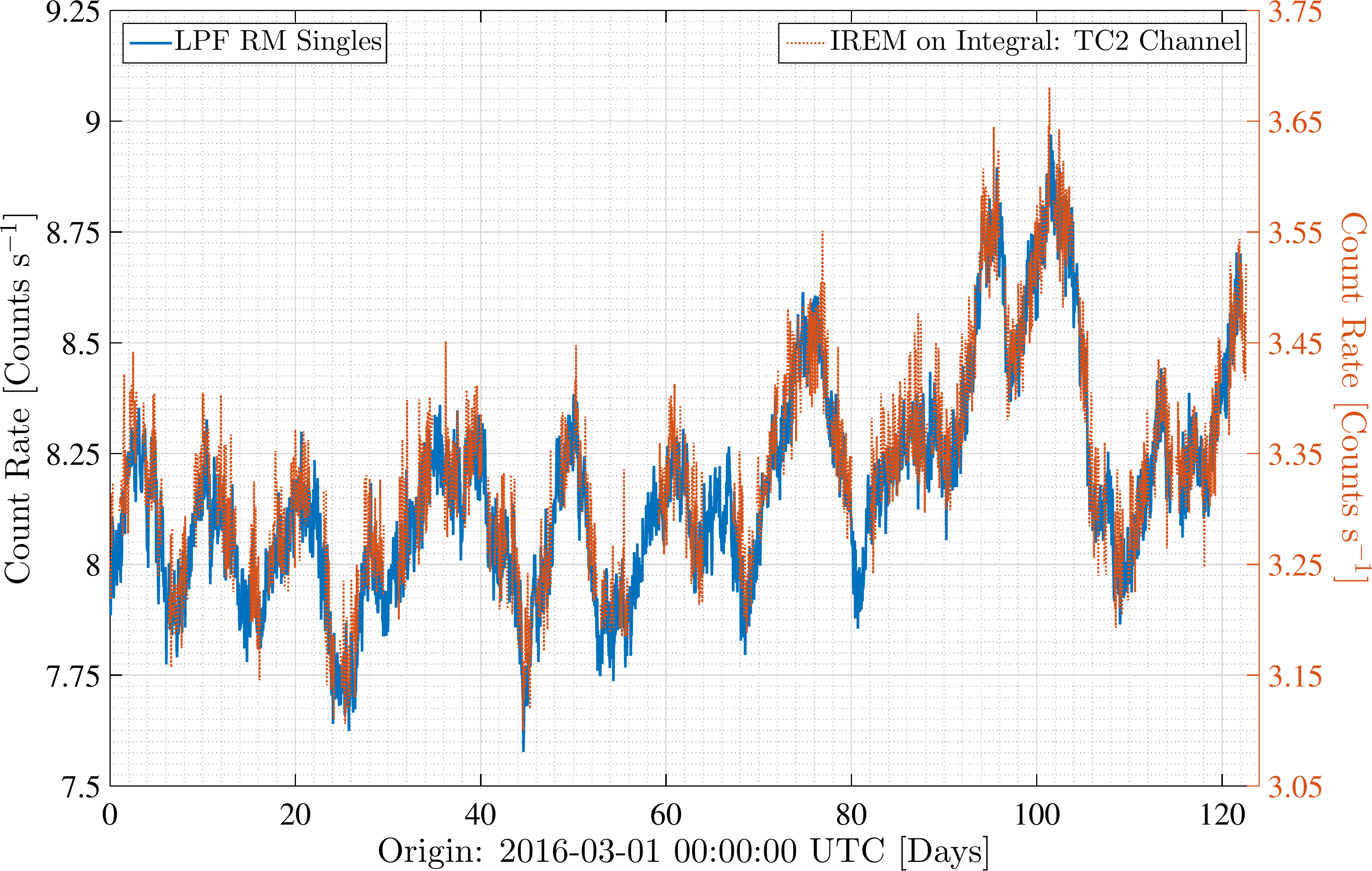}
\caption[IREM Comparison.]{\label{fig:iremComparison} The count rate measured over a four month period by the Pathfinder monitor (left axis) and the IREM TC2 channel (right axis).}
\end{figure}

Indeed, transient variations on such time-scales have been observed in the GCR flux during several solar cycles \cite{Forbush1938, Gupta2009} and to a lesser extent, modulations with 13 to 14 day periods in a variety of solar parameters have also been studied \cite{Mursula1996}. Over the years such variations have been attributed to two possible causes: those related to heliospheric current sheet crossings and those related to co-rotating interaction regions. 

Considering the heliospheric current sheet (HCS) first, as the solar wind propagates out into the solar system it also carries with it the Sun's embedded magnetic field, forming a Archimedean spiral due to the Sun's rotation. Lying within the ecliptic, the HCS is the thin boundary extending out into the solar system where on one side the open magnetic field lines return to the Sun while on the other side they are directed away. Predominately due to the solar magnetic dipole being tilted with respect to the Sun's rotational axis, the HCS becomes warped as it spirals outward and is commonly compared to a twirling ballerina's skirt \cite{Smith2001}. As the HCS co-rotates with the Sun, the Earth passes above and below the sheet where each region can be considered as a separate magnetic sector. When two sectors are present the Earth passes through the HCS boundary twice per solar rotation, but four or even six sectors have previously been detected.

There are a large number of studies that have looked at how the HCS can modulate the received GCR flux, with many focusing on variations around HCS crossings. Such studies tend to combine data from multiple crossings during a particular epoch, so that event specific variation does not obscure general trends. Newkirk et al. \cite{Newkirk1981} used ground based neutron monitors to observe the count rate of $5\,\textup{GeV}$ protons varying with distance (latitude) from the HCS and claimed the count rate falls with increasing latitude from the HCS with data spanning two months each in 1965 and 1975, both during solar minimum. Using a 3-dimensional model, Kota et al. \cite{Kota1982} went on to show that particle drift within the IMF plays an important role in explaining such observations. Using data spanning 1971-1979 and 1981-1990, El-Borie et al. \cite{ElBorie1998} found similar behaviour but that increases in GCR flux were greater during Away-Toward sector crossings than for Toward-Away.

A second possible cause of GCR modulation are co-rotating interaction regions (CIRs). Such regions are produced when a fast solar wind stream (originating from a coronal hole) catches up and compresses a slow solar wind stream, leading to an increase in solar wind particle density as well as magnetic field intensity. Such a region rotates with the Sun and can act to inhibit GCR propagation, decreasing the observed GCR flux as it passes. Richardson et al. \cite{Richardson1996} looked in detail at this phenomenon, studying 305 events between 1973 to 1987 using data from three different spacecraft (IMP 8, Helios 1 and Helios 2). For GCRs above $60\,\textup{MeV}$ they found that almost all CIRs caused depressions (on average by $3.0\,\pm\,1.7\textup{\%}$) with individual streams causing depressions from $0\,\textup{to}\,8\textup{\%}$ and that solar wind speed and depth of the depression were anti-correlated. Note that the oscillation seen in the Pathfinder radiation monitor data is similar in magnitude and is sensitive to a similar energy of incident GCR. It is also possible to have two or more CIRs active at the same time that lead to observed modulation periods less than the expected 26 days were there only one.

\begin{figure*}[]
\centering
\includegraphics[width=1.0\textwidth]{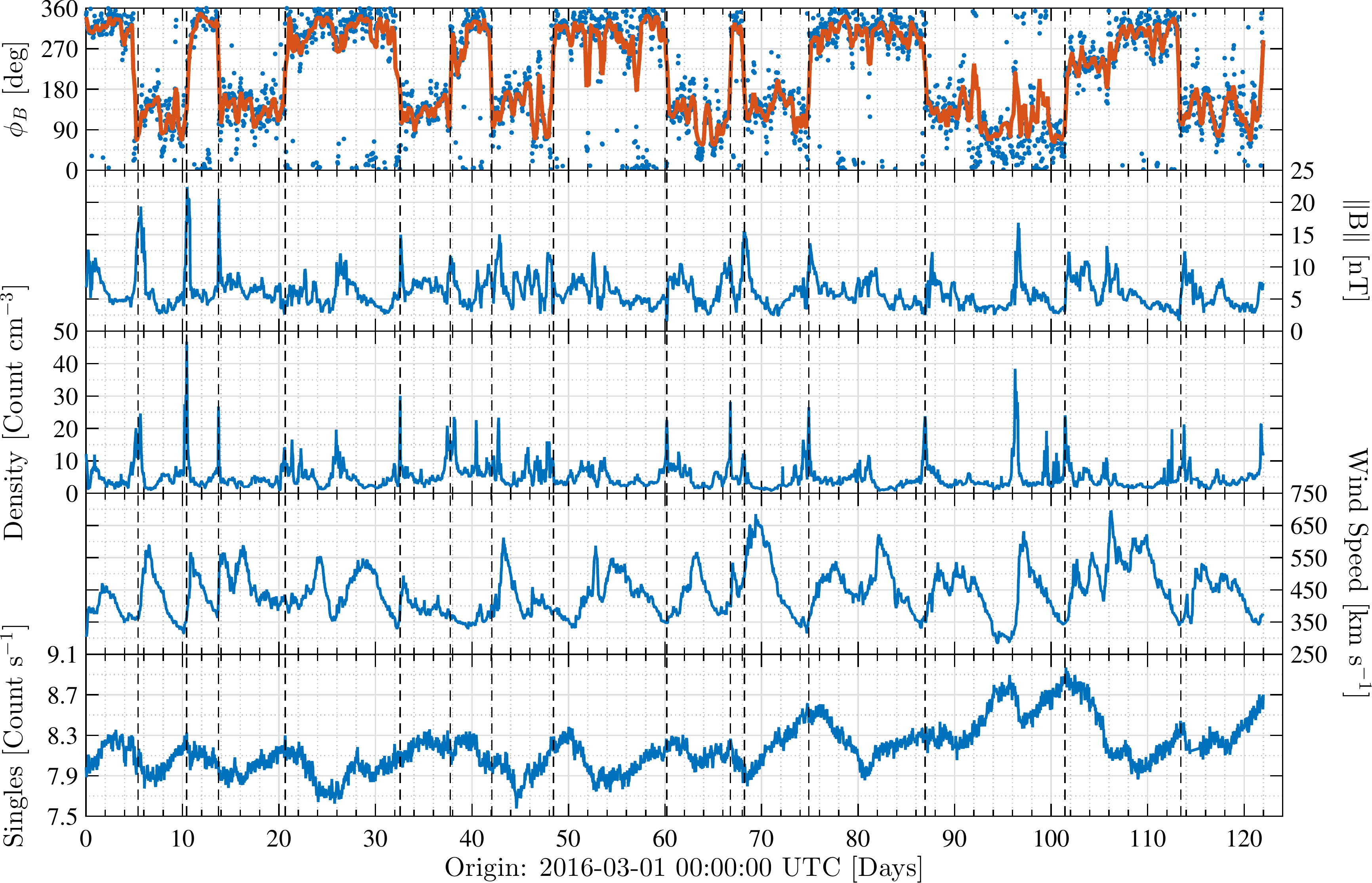}
\caption[Heliospheric Parameters.]{\label{fig:Heliosphere} In descending order from the top panel: The polar angle of the IMF with a smoothed line overlaid, the magnitude of the IMF, the proton density and the solar wind speed all measured by ACE. The bottom panel shows the singles count from the Pathfinder radiation monitor. All data are hourly averaged and the dashed vertical lines show the HCS crossings, as defined by transitions in the polar angle, $\phi_{B}$.}
\end{figure*}

To show that some combination of HCS crossings and CIRs offer a plausible explanation for the observed Pathfinder radiation monitor oscillation we turned to the publicly accessible data from the Advanced Composition Explorer (ACE) \cite{McComas1998}. Launched in 1997 it monitors several properties of the solar wind as well as the IMF via a suite of instruments. Like Pathfinder it is in a Lissajous orbit around L1 with the two spacecraft being separated by between $3\,\textup{to}\,7\times10^{5}\,\textup{km}$ over the course of the Pathfinder mission. As such it offers a useful time-shifted measurement of the solar wind environment relevant to Pathfinder. To confirm that this was indeed the case we compared 1-minute averaged Pathfinder measurements of the IMF with those made by ACE which showed between a $0\,\textup{to}\,30\,\textup{minute}$ time delay between measurements, consistent with solar wind propagation times between the spacecraft depending on their relative positions.

By definition, as one crosses the HCS boundary the dominant direction of the magnetic field is expected to change. To determine when this occurred we used a similar methodology to that described by Thomas et al. \cite{Thomas2014} which involved looking for sudden but prolonged changes in the magnetic polar angle $\phi_{B}$. In the geocentric solar ecliptic (GSE) coordinate system, $\phi_{B}$ is the angle between the magnetic field direction in the $xy$ plane and the positive y-axis, where a value of $0^{\circ}$ points directly toward the Sun and $180^{\circ}$ points away. However, due to it spiralling out from the Sun, the polar angle of the interplanetary magnetic field is rotated by approximately $45^{\circ}$ so that it transitions between average values of $135^{\circ}$ and $315^{\circ}$. Spanning the same four month period as Figure \ref{fig:iremComparison}, Figure \ref{fig:Heliosphere} shows a plot of $\phi_{B}$, where the dashed lines indicate possible crossings. As measured by ACE, the same figure also shows the magnitude of the IMF, the proton density and the solar wind speed, with the final panel showing the Pathfinder radiation monitor singles count for comparison.

Referring to the top panel in Figure \ref{fig:Heliosphere} one can see several periods of the approximately 26-day oscillation (leading edge on days 21, 48, 75 \& 101), presumably due to two dominant sectors of the HCS. Generally, these crossings are also coincident with a peak in the Pathfinder radiation monitor singles count, as suggested by \cite{Newkirk1981} and \cite{ElBorie1998}. On top of this we see three occurrences of a less well defined structure (leading edge on days 10, 38 \& 67). This feature seems to die away as it is not visible where one would expect around day 95, although there is a spike in the other parameters at this time, as well as a pronounced dip in the Pathfinder radiation monitor data. Around HCS crossings we generally, though not on every occasion, see sudden spikes in the amplitude of the magnetic field and proton density, with few spikes being seen outside of these crossing times. The solar wind speed is perhaps the most interesting in that we see a weak anti-correlation with the Pathfinder radiation monitor singles count, seen most clearly in the latter half of the dataset. In addition, there are several instances of sharp changes in the solar wind speed being coincident with sudden dips in the Pathfinder radiation monitor singles count, in particular on day 96. Again, this is what one would expect given the observations described in \cite{Richardson1996}. Ultimately, a detailed understanding of the complex relationship between the time-varying heliosphere and the resulting effect on the GCR flux is beyond the scope of this paper. However, the relatively high count rate of the publicly available Pathfinder dataset is already proving useful for this endeavour \cite{Grimani2017}, with more work in preparation.


\section{Modelling}
\label{sec:modelling}

The GCR proton and helium fluxes are the main inputs into our test mass charging simulations, \cite{Araujo2005, Wass2005}. Having made a variety of test mass charging measurements over the course of the mission we therefore require an estimate of each flux at the time a measurement was made in order to verify the charging simulation. In this section we show how our software model of the radiation monitor, when combined with actual measurements, can be used to estimate the proton and helium fluxes on a given day.

Having been verified by on ground proton beam testing, the GEANT4 part of the radiation monitor model remained unchanged from that described in \cite{Hollington2011}, with one exception. To approximate the spacecraft the simulated monitor was placed inside a $1.42\,\textup{mm}$ thick spherical shell of aluminium, equivalent to the shielding the actual spacecraft provided \cite{S2ASURS2031}. The second part of the software model, which simulates the electronics side of the monitor in MATLAB, was broken down into two parts; the random electronics noise added to every event and the threshold properties that were applied to every event.

\begin{figure}[]
\centering
\includegraphics[width=0.45\textwidth]{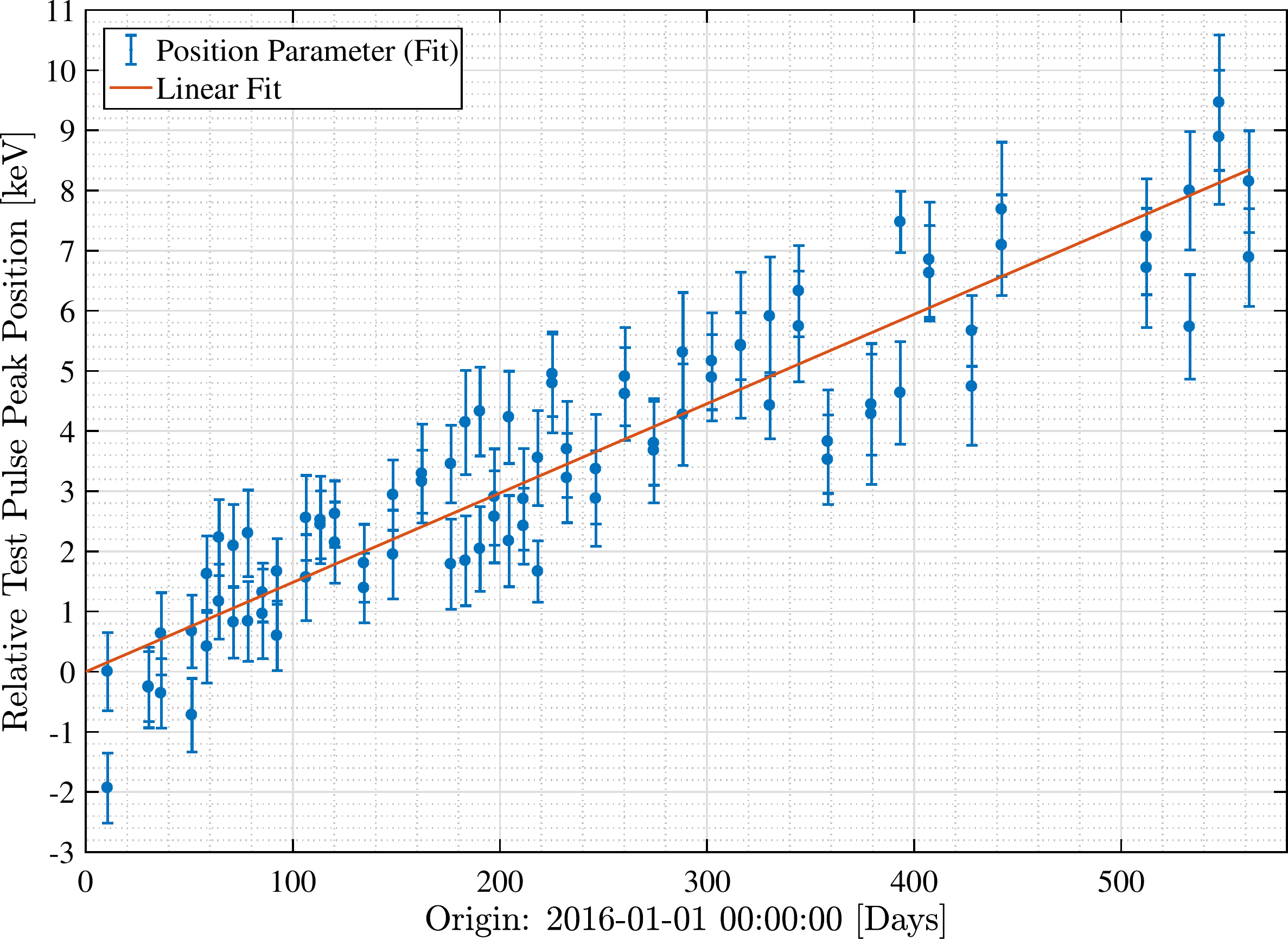}
\includegraphics[width=0.45\textwidth]{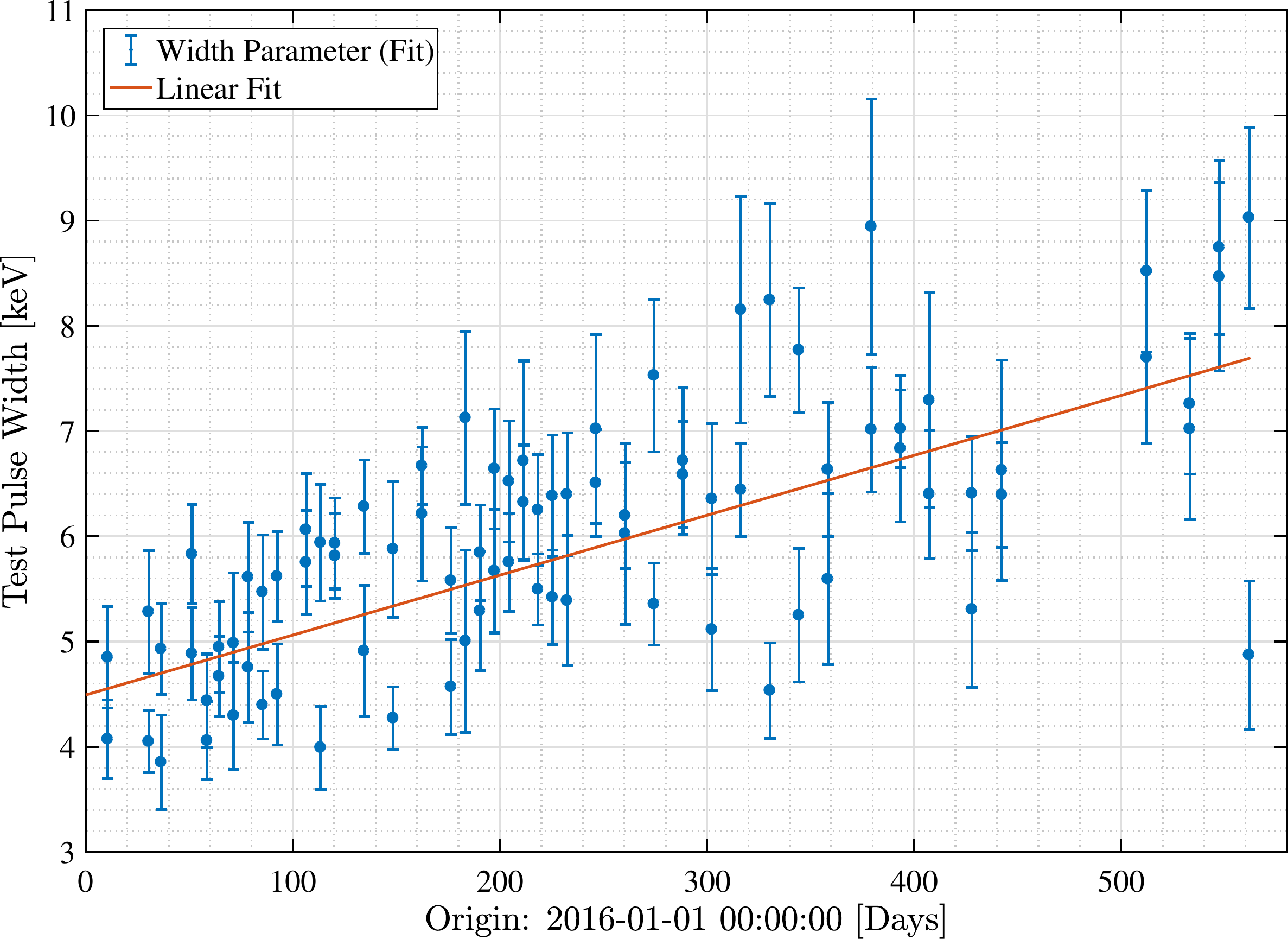}
\caption[Test pulse drift.]{\label{fig:pulseDrift} Top: The relative peak position of the deposited energy test pulses was found to drift linearly over the course of the mission. Note the absolute offset was assumed to be zero at the start of the mission but was in fact unknown. Bottom: The measured width ($\sigma_{noise}$) of the deposited energy pulses also appeared to drift slightly over the course of the mission.}
\end{figure}

As we had done during ground testing, the noise from the electronics was modelled with a Gaussian distribution converted to the equivalent in deposited energy and defined by $\mu_{noise}$ and $\sigma_{noise}$, both in units of keV. For every event that deposited energy in a diode this distribution was randomly sampled and added independently as noise to each simulated deposit. Over the course of the mission a series of test pulses were injected every one to two weeks, with each pulse being recorded in the deposited energy spectrum. Given any noise coming from the electronics should have been added to these pulses, they offered a way of assessing the electronics noise as well as its stability in time. Reassuringly, the pulses in the deposited energy spectrum appeared approximately Gaussian and a simple least squares fit allowed a $\mu_{noise}$ and $\sigma_{noise}$ to be obtained for each test pulse. The resulting time-series are shown in Figure \ref{fig:pulseDrift}.

Both the relative position and width of the test pulses appeared to drift up over the mission, with both effects being attributed to the electronics noise ($\mu_{noise}$ and $\sigma_{noise}$) rather than the generation of the pulses themselves. This assumption is supported by the fact that the peak in the deposited energy spectrum also drifted by the same amount over the mission. Unfortunately the expected absolute position of the test pulses was unknown and therefore $\mu_{noise}=0\,\textup{keV}$ was assumed as the initial value with the relative drift quantified thereafter. Reassuringly, the initial $\sigma_{noise}=4.5\,\textup{keV}$ was consistent with an expected value of $4.24\,\textup{keV}$ \cite{S2IFAEDDD3002}. Although both drifts were small (considering the deposited energy bin widths are $\approx4.88\,\textup{keV}$) a time dependant correction was added to our model of the electronics noise.

The second part of our electronics model concerned the two diode's independent commandable threshold levels, which determined if a particular deposited energy was recorded as a hit or not. These were also modelled as having a Gaussian distributed noise, with $\mu_{thresh}$ taking the nominally commanded threshold level, which was $60.7\,\textup{keV}$ for all the data considered here. Based on calibration measurements made early in the mission, the threshold noise appeared to be comparable to the electronics noise and therefore $\sigma_{thresh}=\sigma_{noise}$ was assumed for the analysis presented here.

\begin{figure}[]
\centering
\includegraphics[width=0.45\textwidth]{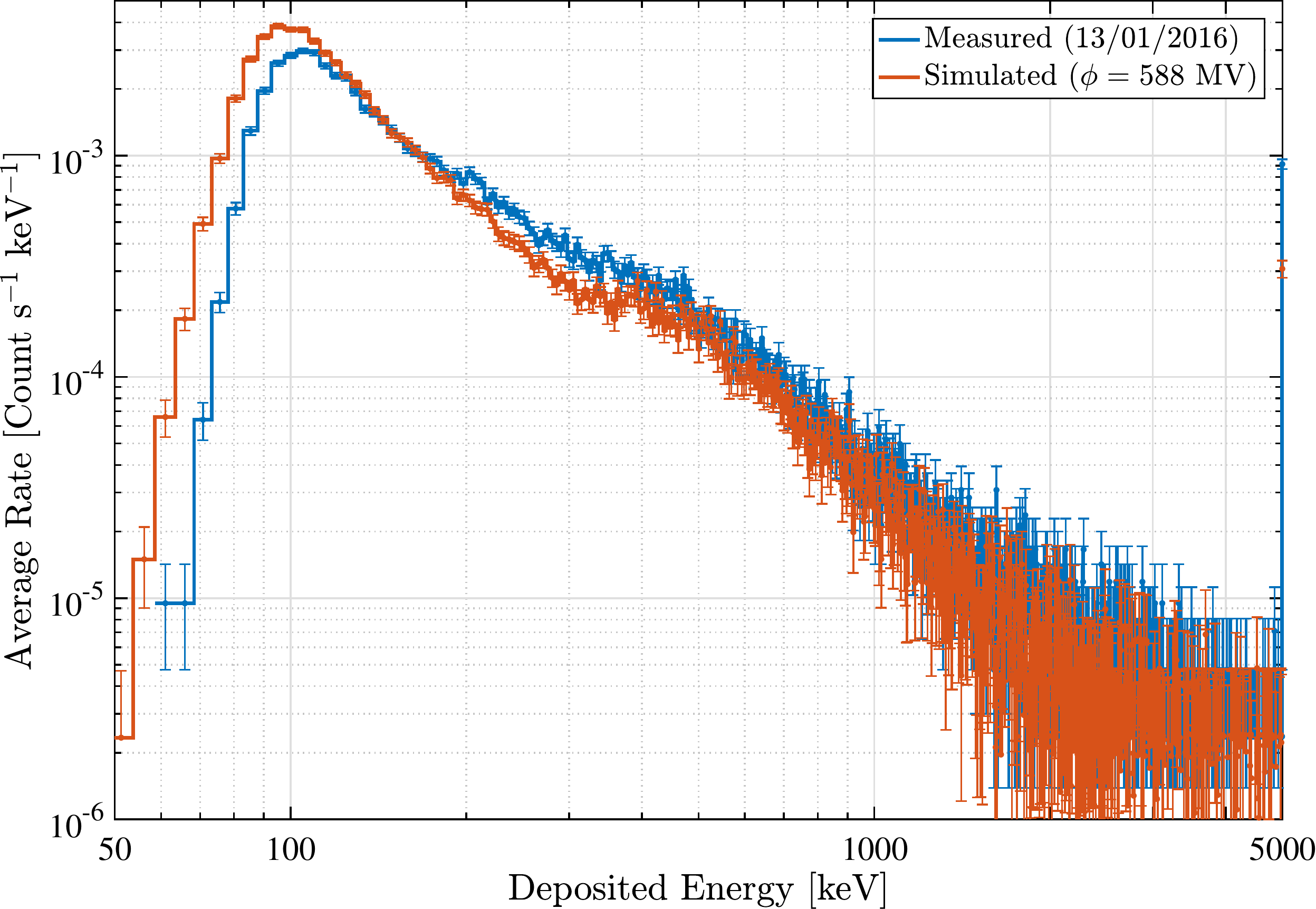}
\caption[PhiFitting.]{\label{fig:phiFits} A comparison between a measured deposited energy spectrum (averaged over an entire day) and the best fit from simulation.}
\end{figure}

We parametrised the differential incident energy spectrum for both protons and helium using $\phi$, as described in Sec. \ref{sec:LIS}. This gave us a complete description for our radiation monitor model; $\phi$ describes the proton and helium GCR spectra that act as the input to the GEANT4 model, the GEANT4 model outputs a list of deposited energies in each diode given the input GCR spectra, a randomly sampled electronics noise is independently added to each diode for every event, and finally each diode is tested against the randomly sampled detection thresholds. For a particular input $\phi$, the complete model returns both a simulated deposited energy spectrum in the back diode as well as a simulated singles count from both diodes. By using a measured deposited energy spectrum we were therefore able to fit for the free parameter $\phi$, with an example of such a fit for the first full day of data is shown in Figure \ref{fig:phiFits}.

As one can see in the example from the $13^{\textup{th}}$ January 2016, the fit for $\phi$ ends up compromising in different sections of the spectrum in order to minimise the overall residual. The simulated spectrum overestimates the peak count by around  20\% but underestimates by about 20\% between $200\,\textup{keV}$ and $400\,\textup{keV}$. Above this it systematically underestimates by roughly 10\%, though the data are noisy due to relatively few counts in these bins. Visually there appears to be a significant discrepancy at the low energy edge, but this is accentuated due to the logarithmic scale on the x-axis with the spectra only actually offset by around $1.5\,\textup{bins}$ equivalent to $8\,\textup{keV}$.

There are potentially a large number of parameters that could be tuned to improve the quality of the fit and indeed this was initially attempted. For example, there are several ways of varying the position of the edge between $50\,\textup{keV}$ and $100\,\textup{keV}$. Just considering the electronics, increasing the mean diode noise by $+8\,\textup{keV}$ would lead to a shift in the deposited spectrum that could account for the discrepancy. Alternatively the nominal $60.7\,\textup{keV}$ threshold level could have had an offset of $\approx+30\,\textup{keV}$ that would cut this edge. Both are feasible but disentangling each contribution proved difficult, even before one considers that each diode could potentially have different noise and/or threshold properties. Given the level of agreement with the nominal parameter values we ultimately decided against improving the fits further and risk over-fitting, particularly as some discrepancies may lie with imperfections in the modelled GCR spectrum, detector geometry description or GEANT4 physics modelling.

Using the same model and the $\phi$ obtained from fitting to the spectral data, we also generated a simulated singles count. We found this to be systematically lower that the measured singles count by approximately $0.7\,\textup{counts s}^{-1}$ or around 7\% of the total. This difference is partially explained by noise occasionally triggering the detector as although our model adds random noise to every simulated deposited energy event it is incapable of generating a singles event due purely to noise. Dark singles count rates of $\approx0.2\,\textup{counts s}^{-1}$ were measured on ground \cite{S2IECTR3114}. Even ignoring the possibility that the dark count has since increased, the remaining discrepancy is at about the 5\% level which we deemed acceptable.

\begin{figure}[]
\centering
\includegraphics[width=0.45\textwidth]{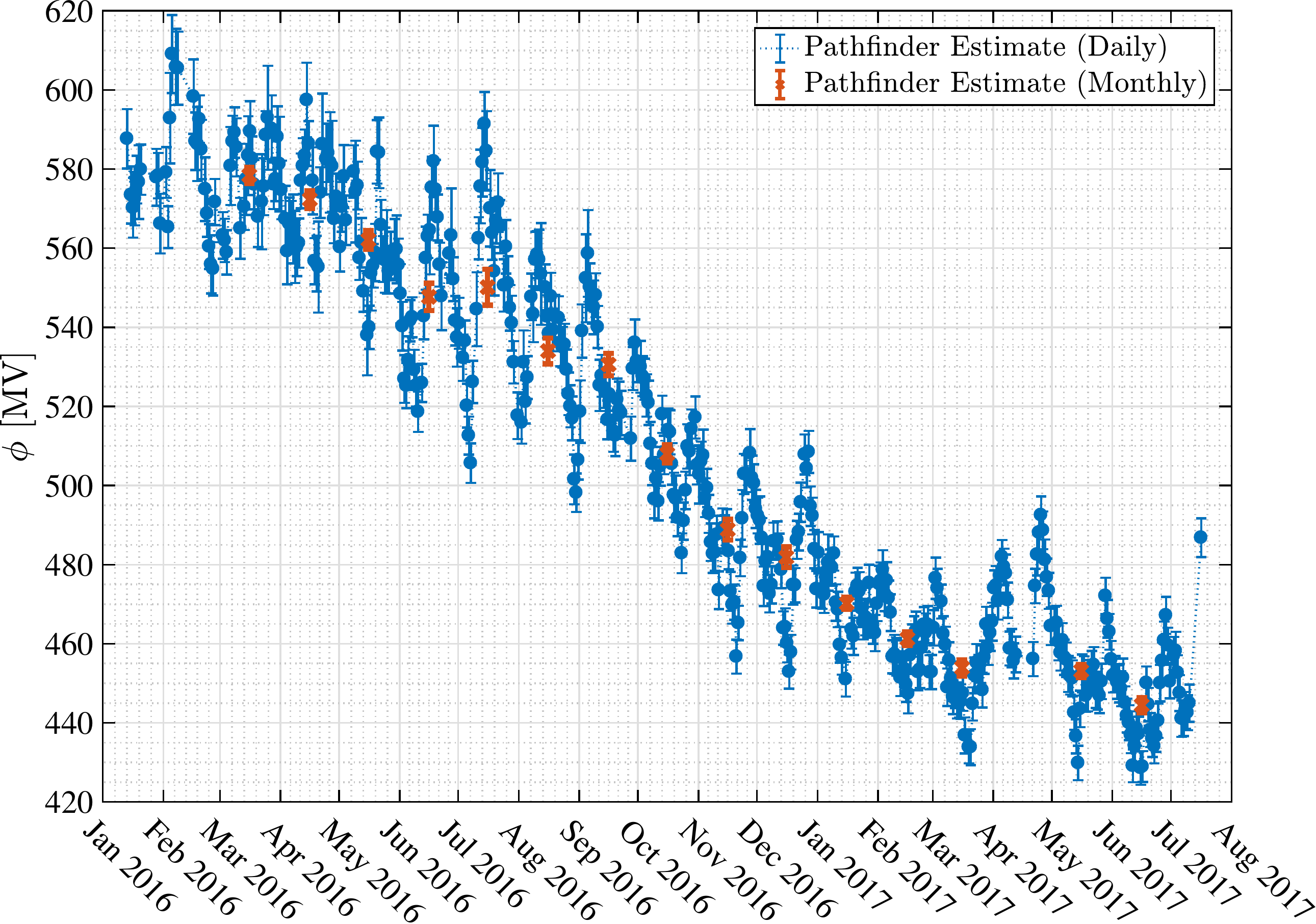}
\caption[PhiEstimates.]{\label{fig:Phiestimates} $\phi$ time-series if one fits for everyday of available mission data. Note, the time-series shows the same 13 and 26 day fluctuation in the GCR flux as discussed in Sec. \ref{sec:GCRmodulation}. Also shown are monthly averages for months with no data gaps.}
\end{figure}

\begin{figure}[]
\centering
\includegraphics[width=0.45\textwidth]{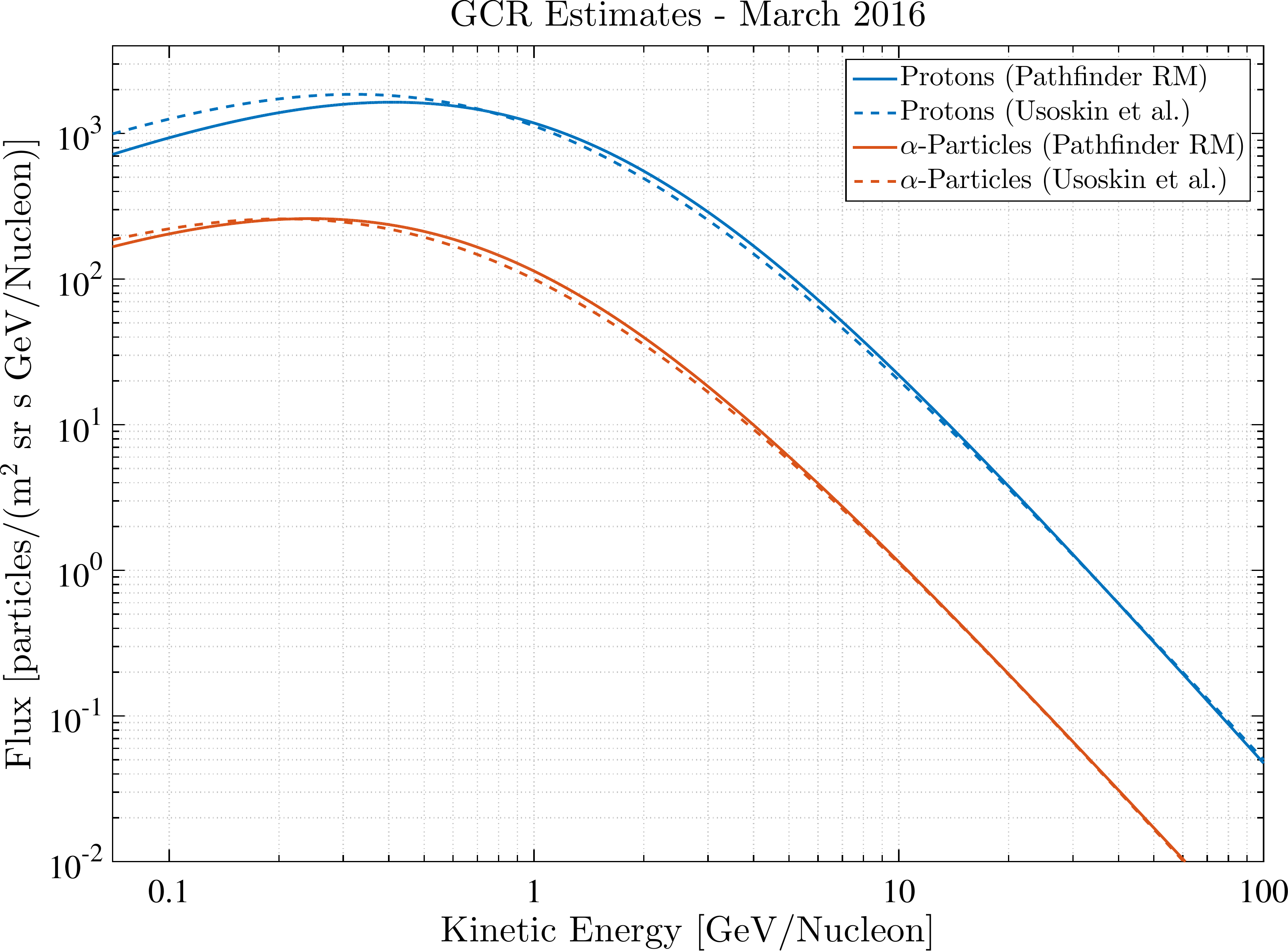}
\includegraphics[width=0.45\textwidth]{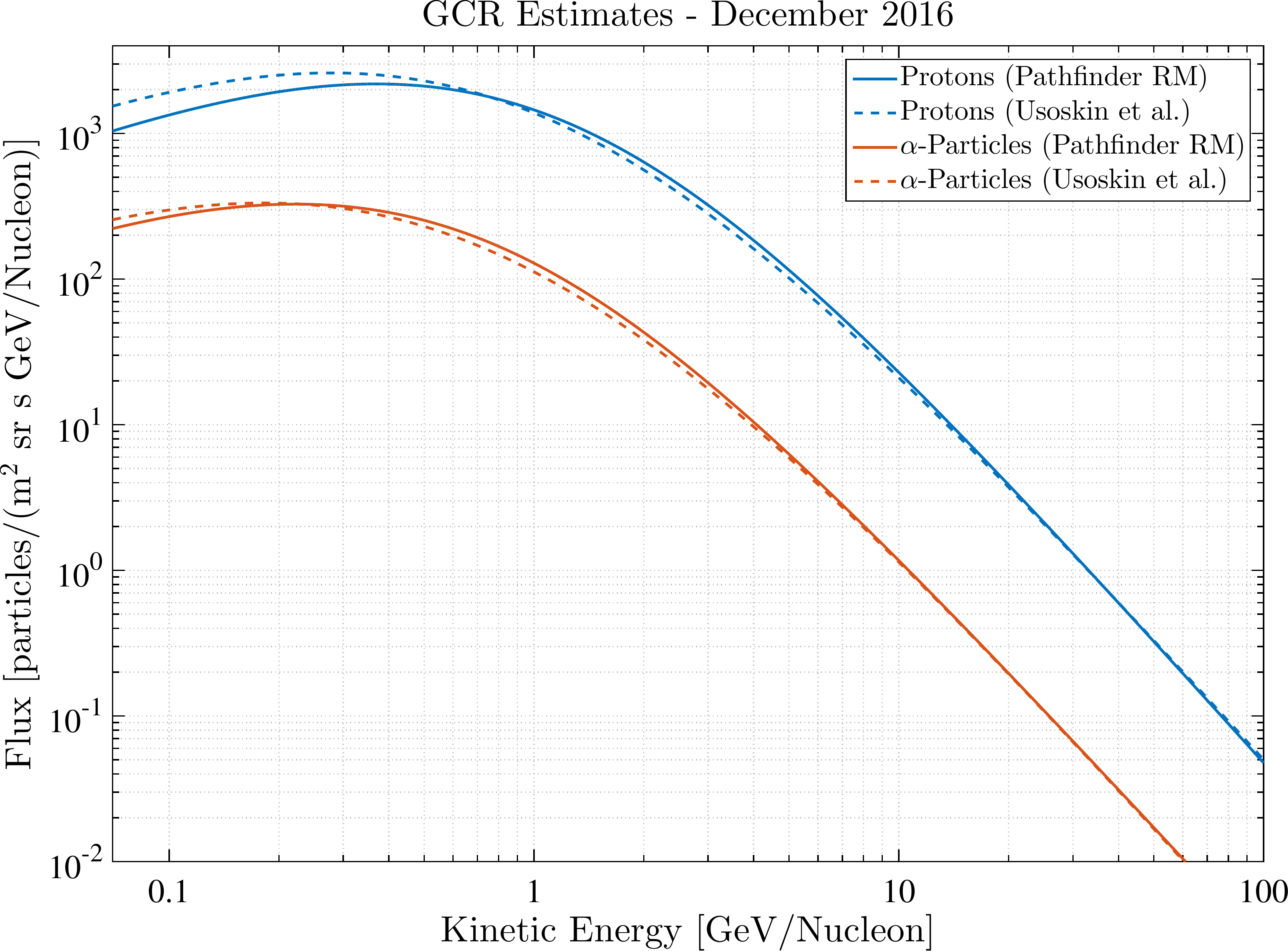}
\caption[GCR Estimates.]{\label{fig:GCRestimates} Top: The estimated differential energy spectra for both protons and $\alpha$-particles, averaged over the month of March 2016. The solid lines are obtained from the Pathfinder analysis while the dashed lines are based on that of Usoskin et al. \cite{Usoskin2005, Usoskin2011}. Bottom: The equivalent estimates for the month of December 2016. Note the change from March 2016 are most pronounced at low energies.}
\end{figure}

Figure \ref{fig:Phiestimates} shows the resulting $\phi$ time-series if one fits for every day of the mission. To assess our estimates of $\phi$ we compared it to monthly estimates published at \url{http://cosmicrays.oulu.fi/phi/Phi_mon.txt} by the world neutron monitor network using an analysis described in \cite{Usoskin2005, Usoskin2011}. They use an older expression for the local interstellar spectrum which prevents direct comparison between our values for $\phi$. However, we can compare the resulting differential energy spectra predicted by both models with Figure \ref{fig:GCRestimates} showing those for March 2016 and December 2016, the first and last dates when predictions exist from both.

While there is good general agreement there are clear differences in the two estimates, particularly for the number of protons with energies less than $500\,\textup{MeV}$. Overall however, we find good agreement in the integrated flux above $70\,\textup{MeV}$ between the two analyses. During March 2016 we estimate $0.329\pm0.001$ protons cm$^{-2}$ sr$^{-1}$ s$^{-1}$ while Usoskin et al. estimate $0.322\pm0.018$ protons cm$^{-2}$ sr$^{-1}$ s$^{-1}$. For December 2016 estimates of $0.397\pm0.003$ protons cm$^{-2}$ sr$^{-1}$ s$^{-1}$ and $0.394\pm0.025$ protons cm$^{-2}$ sr$^{-1}$ s$^{-1}$ were obtained respectively.


\section{Conclusions}

The Pathfinder radiation monitor successfully measured the interplanetary charged-particle environment at L1, from January 2016 until mission completion in July 2017. For particles with incident energies above $\approx70\,\textup{MeV}$ it measured a singles count as well as the energy deposited by coincident events. This provided a diagnostic for the primary differential acceleration measurement but more importantly an \textit{in situ} measure of the particles responsible for test mass charging. Comparing the radiation monitor data with our direct test mass charging measurements will help us to verify our current charging models and enable a more realistic estimate for the charge induced noise for LISA.

Given its relatively high counting statistics compared to similar channels on other detectors, the Pathfinder radiation monitor also offered an opportune measurement of short term GCR flux variability. Changes in the GCR flux on time scales from 13 to 26 days were observed over the mission, ultimately due to changes in the heliosphere. Similar large scale fluctuations in the heliosphere may lead to correlated but time delayed changes in charging rate for the test masses aboard the distant spacecraft of the LISA constellation.

LISA will likely also experience SEP events that will temporarily cause significant increases in test mass charging rates. Although several SEP events occurred during the Pathfinder mission, none were above the $\approx100\,\textup{MeV}$ energy threshold required to penetrate the sensor and enhance charging rates. However, the largest such event did produce an observable Forbush decrease which may yet prove useful for future analyses.

Using a detailed two part model, we have also shown how the energy deposited by coincident events can be used to obtain an estimate of the incident cosmic ray spectra. In terms of integrated flux, our estimates were shown to be in good agreement with those produced by neutron monitoring networks on ground. To achieve this linear drifts in the electronics noise, identified by the regularly injected test pulses, had to be accounted for. While these drifts were relatively small, had they continued they would have begun to limit data quality after several years of operation. Such electronics drift will need to be rectified in a potential LISA radiation monitor design given a possible multi-year mission lifetime.

Forming the basis for a future paper, our estimates of the incident cosmic ray spectra will provide the main input to our test mass charging models which will then be compared to direct measurements. Such modelling will fundamentally improve our understanding of test mass charging and ultimately aid the design of the LISA mission.


\section*{Acknowledgements}

We acknowledge the NMDB database (\url{www.nmdb.eu}), founded under the European Union's FP7 programme (contract no. 213007) for providing data. In addition, the Oulu NM station is operated by the University of Oulu with data available at \url{http://cosmicrays.oulu.fi}. We thank the ACE SWEPAM instrument team and the ACE Science Centre for providing the ACE data. The GOES data mentioned in this study can be obtained from the National Geophysical Data Centre at \url{http://satdat.ngdc.noaa.gov/sem/goes/data/new_avg/}. The full Pathfinder radiation monitor dataset can be found at \url{http://lpf.esac.esa.int/lpfsa/}.

This work has been made possible by the LISA Pathfinder mission, which is part of the space-science program of the European Space Agency. The French contribution has been supported by CNES (Accord Specific de Projet No. CNES 1316634/CNRS 103747), the CNRS, the Observatoire de Paris and the University Paris-Diderot. E.~Plagnol and H.~Inchausp\'e would also like to acknowledge the financial support of the UnivEarthS Labex program at Sorbonne Paris Cit (Grants No. ANR-10-LABX-0023 and No. ANR-11-IDEX-0005-02). The Albert-Einstein-Institut acknowledges the support of the German Space Agency, DLR. The work is supported by the Federal Ministry for Economic Affairs and Energy based on a resolution of the German Bundestag (Grants No. FKZ 50OQ0501 and No. FKZ 50OQ1601). The Italian contribution has been supported by Agenzia Spaziale Italiana and Instituto Nazionale di Fisica Nucleare. The Spanish contribution has been supported by Contracts No. AYA2010-15709 (MICINN), No. ESP2013-47637-P, and No. ESP2015-67234-P (MINECO). M.~Nofrarias acknowledges support from Fundacion General CSIC Programa ComFuturo). F.~Rivas acknowledges support from a Formacin de Personal Investigador (MINECO) contract. The Swiss contribution acknowledges the support of the Swiss Space Office (SSO) via the PRODEX Programme of ESA. L.~Ferraioli acknowledges the support of the Swiss National Science Foundation. The UK groups wish to acknowledge support from the United Kingdom Space Agency (UKSA), the University of Glasgow, the University of Birmingham, Imperial College London, and the Scottish Universities Physics Alliance (SUPA). N.~Korsakova would like to acknowledge the support of the Newton International Fellowship from the Royal Society. J.\,I.~Thorpe and J.~Slutsky acknowledge the support of the U.S. National Aeronautics and Space Administration (NASA).



\bibliographystyle{unsrt}
\bibliography{bibliography}


\end{document}